%% file: Project.tex
\pdfoutput=1

\documentclass[aps,pre,twocolumn]{revtex4-1}
\usepackage{hyperref}

\usepackage[utf8]{inputenc}
\usepackage[russian,english]{babel}
\usepackage{bm}
\usepackage{amssymb}
\usepackage{amsxtra}
\usepackage[mathscr]{eucal}
\usepackage{xcolor}
\usepackage{graphicx}

\definecolor{mybrown}{rgb}{0.5,0,0}
\definecolor{mylightgreen}{rgb}{0,0.83,0}
\definecolor{mygreen}{rgb}{0,0.5,0}

\usepackage{mathtools}
\usepackage{chngcntr}
\usepackage{etoolbox}

\input def

\begin{document}

\selectlanguage{english}
%\title{Steady-state kinetic temperature distribution in a
%two-dimensional square harmonic scalar lattice lying in a viscous environment
%and subjected to a point heat source\thanks{
%This work is supported by Russian Science Foundation (Grant No. 18-11-00201).}}
%\author{Serge N.~Gavrilov \and Anton M.~Krivtsov}
%\date{\GITDate\quad commit:\ \GITHash}
%
%\titlerunning{Steady-state kinetic temperature distribution}
%\institute{
%S.N.~Gavrilov \at
%Institute for Problems in Mechanical Engineering RAS, V.O., Bolshoy pr.~61,
%St.~Petersburg, 199178, Russia \\
%\email{serge@pdmi.ras.ru}           
%\and
%S.N.~Gavrilov \at
%Peter the Great St.~Petersburg Polytechnic University (SPbPU),
%Polytechnicheskaya str.~29, St.Petersburg, 195251, Russia
%\and
%A.M.~Krivtsov \at
%Peter the Great St.~Petersburg Polytechnic University (SPbPU),
%Polytechnicheskaya str.~29, St.Petersburg, 195251, Russia\\
%\email{akrivtsov@bk.ru}
%\and
%A.M.~Krivtsov \at
%Institute for Problems in Mechanical Engineering RAS, V.O., Bolshoy pr.~61,
%St.~Petersburg, 199178, Russia}
%
%\selectlanguage{english}
\title{Thermal equilibration in a one-dimensional damped harmonic crystal}
\author{S.N.~Gavrilov}
\email{serge@pdmi.ras.ru}
\affiliation{Institute for Problems in Mechanical Engineering RAS, V.O., Bolshoy
pr.~61, St.~Petersburg, 199178, Russia}
\affiliation{Peter the Great St.~Petersburg Polytechnic University (SPbPU),
Polytechnicheskaya str.~29, St.Petersburg, 195251, Russia}
\author{A.M.~Krivtsov}
\email{akrivtsov@bk.ru}
\affiliation{Institute for Problems in Mechanical Engineering RAS, V.O., Bolshoy
pr.~61, St.~Petersburg, 199178, Russia}
\affiliation{Peter the Great St.~Petersburg Polytechnic University (SPbPU),
Polytechnicheskaya str.~29, St.Petersburg, 195251, Russia}

\begin{abstract}
The features for the unsteady process of thermal equilibration (``the fast
motions'') in
a one-dimensional harmonic crystal 
%with nearest-neighbor interactions 
lying in a
viscous environment (e.g., a gas) are under investigation. It is assumed that
initially the displacements of all the particles are zero and the particle
velocities are random quantities with zero mean and a constant variance, thus,
the system is far away from the thermal equilibrium.  It is known that in the framework of
the corresponding conservative problem the kinetic and potential energies
oscillate and approach the equilibrium value that equals a half of the initial
value of the kinetic energy. We show that the presence of the external
damping qualitatively changes the features of this process. 
%In the latter
%case, the limiting values for the kinetic and potential energies, clearly, are zero. 
The unsteady process generally has two stages. At the first stage
oscillations of kinetic and
potential energies with decreasing amplitude, subjected to exponential decay, can be observed (this stage
exists only in the underdamped case). At the second
stage (which always exists), 
the oscillations vanish, and the energies
are subjected to a power decay. The large-time asymptotics for the energy is proportional to $t^{-3/2}$ in the case
of the potential energy and to $t^{-5/2}$ in the case the kinetic energy.
Hence, at large values of time the total energy of the crystal is mostly the potential
energy. The obtained analytic results are verified by independent numerical
calculations.
\end{abstract}

\maketitle

\input{def-fast}

\input{fast}

\selectlanguage{english}

\end{document}

%% file: def.tex
\def\I{\mathrm{i}}

\def\F#1#2#3{#1_{{\mathscr F_{#3}^#2}}}

\def\be#1{\begin{equation}\label{#1}}
\def\ee{\end{equation}}
\newcommand {\ba}[2]{\be{#1}\begin{array}{#2}}
\newcommand {\ea}{\end{array} \ee}

\newcommand{\q}{\,,\quad}

\renewcommand{\=}{\stackrel{\mbox{\scriptsize def}}{=}}

\def\({\left(}
\def\){\right)}
\def\av#1{\langle{#1}\rangle}

\let\w = \omega

\def\dt{\partial_t}

\def\LL{{\mathfrak L}}

\let\de = \delta
\let\la = \Lambda
\def\L{{\mathcal L}}

\def\d{{\cal D}}

\let\ka=\kappa

\def\sign{\operatorname{sign}}

\def\NEW#1{{\color{black}#1}}

\def\LL{\NEW{\omega_0^2\L}}

\def\d{\mathrm d}
\def\DDD{\mathrm d}

\def\Im{\operatorname{Im}}

\def\kao{\kappa_{ii}(0)}
\def\PV{\operatorname{PV}}
\def\mkE{\bar{\mathscr E}_0}
\def\Ci{\operatorname{Ci}}
\def\Chi{\operatorname{Chi}}
\def\q{{\mathfrak q}}

%% file: def-fast.tex
\def\F#1#2#3{#1_{{F}}}
\let\d=\DDD
\def\kao{\kappa_{i}(0)}

%% file: fast.tex
%\tableofcontents

\section{Introduction}

In this paper, we study the influence of an external viscous environment
(e.g., a gas) on the unsteady process of thermal equilibration in an infinite
one-dimensional harmonic crystal with nearest-neighbor interactions. 
The model of a damped harmonic crystal 
was used in our recent papers \cite{gavrilov2018heat,gavrilov2019steady},
where we discuss the 
ballistic heat propagation in such a structure (``the slow motions''). As opposed
to \cite{gavrilov2018heat,gavrilov2019steady} now we consider the fast
motions, i.e. the fast vanishing oscillations of the kinetic and potential
energy. This
oscillations are well known for those who deal with molecular dynamics
simulation (see, e.g., \cite{allen2017computer}, Fig.~5.11). 
%In the case under
%consideration in the paper these
%oscillations are caused by random non-equilibrium initial conditions with
%zero mean and a spatially uniform variance. 

We assume that initially the displacements of the crystal particles are zero and
the particle velocities are random quantities with zero mean and a constant
variance. The kinetic energy per particle (as well as the corresponding kinetic
temperature) is distributed spatially uniform, whereas the potential energy
is zero.  Thus, the thermodynamic system is far away from thermal equilibrium. 
It is well known that in the framework of the corresponding conservative
problem (in the absence of external damping)
the kinetic and potential energies oscillate and approach the
equilibrium value that equals a half of the initial value of the total
energy (equipartition of kinetic and potential energy).
For a harmonic crystal the process of thermal equilibration was first time
investigated by Klein \& Prigogine in
\cite{klein1953mecanique}. Thermal equilibration in harmonic
crystals in the conservative case was considered in many studies, e.g.,
\cite{linn1984thermal,hemmer1959dynamic,huerta1971exact,Robertson1969,Robertson1970,kannan2012nonequilibrium,lepri2008stochastic,rieder1967properties,lepri2010nonequilibrium,krivtsov2014energy,babenkov2016energy,kuzkin2017high,guzev2018,Kuzkin-Krivtsov-accepted,kuzkin2017analytical}.
The more complete bibliography can be found in recent paper by
Kuzkin~\cite{kuzkin2019thermal}, where the analytic solution in the integral
form is obtained for an infinite harmonic crystal with an arbitrary
Bravais lattice and a polyatomic cell with an arbitrary structure. The time
evolution the kinetic temperature during the thermal equilibration in a harmonic crystal
was considered in Refs.~\cite{klein1953mecanique,hemmer1959dynamic,krivtsov2014energy,babenkov2016energy,kuzkin2017high,guzev2018,Kuzkin-Krivtsov-accepted,kuzkin2017analytical,kuzkin2019thermal},
the entropy was under consideration in 
Refs.~\cite{huerta1971exact,Robertson1969,Robertson1970,PhysRevE.99.042107}.
Thermal
equilibration for a system of quantum oscillators is considered
in~\cite{Devi2009}.

In the paper, we show that the process of thermal equilibration in the presence of
a viscous external environment is more complicated than in the conservative case
and has two stages in the underdamped case
\footnote{%
In the underdamped case the specific viscosity of the external environment is
small enough such that restriction \eqref{FAST-eta-restriction} is fulfilled},
which is generally assumed in the paper.
In the presence of damping, the limiting values for the kinetic
and potential energies, clearly, are zero.  At the first stage, the {\it
qualitative} description of the process is as follows: the kinetic and
potential energies oscillate approaching an exponentially-decaying curvilinear asymptote.
Unexpectedly, for any positive value of the specific viscosity for
environment,
there is the second stage, which can observed in the underdamped
case only for very large values of time.  
The kinetic and the potential
energies at the second stage are subjected to a power decay.
Another one unexpected result is as follows:
the principal term of the large-time asymptotics  
is proportional to $t^{-3/2}$ in the case of 
potential energy and to $t^{-5/2}$ in the case of the kinetic energy. 
Hence, at very large times the total energy of the harmonic crystal is mostly the potential
energy.

The paper is organized as follows. In Section~\ref{FAST-Sec-formulation}, we
consider the formulation of the problem. In Section~\ref{FAST-SSec-notation},
some general notation is introduced. In Section~\ref{FAST-SSec-stochastic}, we state the basic
equations for the crystal particles in the form of a system of 
ordinary differential equations with random initial conditions.
In Section~\ref{FAST-SSec-covariance}, we
introduce and deal with infinite set of covariance variables. These  
are the mutual covariances of the particle velocities and the
displacements for all pairs of particles. 
We obtain two infinite systems of
differential-difference equations involving only the covariances for {the
particle velocities}, and only the covariances for {the displacements},
respectively. The similar approach was used in previous papers 
\cite{krivtsov2014energy,krivtsov2015heat,
krivtsov-da70,
sokolov2017localized,krivtsov2018one,
babenkov2016energy,
kuzkin2017analytical,
kuzkin2017high,
Kuzkin-Krivtsov-accepted,
murachev2018thermal,
PhysRevE.99.042107,
gavrilov2018heat,gavrilov2019steady}.
In Section~\ref{FAST-sec-uniform} we simplify the obtained equations using the
assumptions of uniformity for the variance of the initial values of the particle
velocities. Finally, we obtain four infinite systems of ordinary differential 
equations for the energetic quantities,
which we call 
the generalized kinetic energy,
the generalized potential energy,
the generalized total energy, 
the generalized Lagrangian.
The 
kinetic energy,
the potential energy,
the total energy, 
and the Lagrangian, are particular cases of those quantities. It is sufficient to
solve the equations for any two of these four energetic quantities to
calculate all of them. We choose the generalized Lagrangian and the
generalized potential energy as basic variables, 
since the corresponding equations have a simpler structure.
In Section~\ref{FAST-Sec-analyt-intform} we use discrete-time Fourier
transform to get the analytical solutions for the Lagrangian and the potential
energy in the integral form.  
In Section~\ref{FAST-section-AS} all the energetic quantities are evaluated
for the large values of time. To do this we use the method of stationary phase and the
Laplace method
\cite{Fedoruk-Saddle}. The corresponding calculations are given in 
Appendices~\ref{FAST-App-L2}--\ref{FAST-App-P0}.
In Section~\ref{FAST-sec-numerics}, we present the
results of the numerical solution of the initial value problem for the system
of ordinary differential equations with random initial conditions 
and compare them with the obtained analytical
solution in the integral and the asymptotic forms.
In Section~\ref{FAST-Sec-discussion}
we discuss the large-time behavior of energetic quantities in
the underdamped case.
Finally, in Section~\ref{FAST-Sec-conclusion}, we discuss the basic
results of the paper.

\section{Mathematical formulation}
\label{FAST-Sec-formulation}
\subsection{Notation}
\label{FAST-SSec-notation}
In the paper, we use the following general notation:
\begin{description}	
\item[$t$] 
%is 
the time;
\item[$H(\cdot)$] 
%is 
the Heaviside function;
\item[$\delta(\cdot)$] 
%is 
the Dirac delta function;
\item[$\langle\cdot\rangle$]
%is
the expected value for a random quantity;
\item[$\de_{pq}$]
%is 
the Kronecker delta ($\de_{pq}= 1$ if  $p=q$, and $\de_{pq} = 0$
otherwise);
\item[$\delta_n$] 
%is such that  $\de_n = 1$ if $n=0$ and $\de_n=0$ otherwise;
$\delta_n\=\delta_{n0}$;
\item[$J_0(\cdot)$] 
%is 
the  Bessel function of the first kind of zero order
\cite{abramowitz1972handbook};
%\item[$I_0(\cdot)$]  the modified Bessel function of the first kind of zero order
%\cite{abramowitz1972handbook};
%\item[$K_0(\cdot)$]  the Macdonald function (the modified Bessel function of the second
%kind) of zero order
%\cite{abramowitz1972handbook};
%\item[$\erfc(\cdot)$]  the complementary error function
%\cite{abramowitz1972handbook};
\item[$\Gamma(\cdot)$] 
%is 
the Euler integral of the second kind (the Gamma
function)
\cite{abramowitz1972handbook};
\item[$C^\infty$] 
%is 
the set of all infinitely differentiable functions;
\item[$\mathbb Z$] 
the set of all integers.
\end{description}

\subsection{Dynamic equations for a crystal and random initial conditions}
\label{FAST-SSec-stochastic}

\begin{figure}[htp]
\centering\includegraphics[width=0.95\columnwidth]{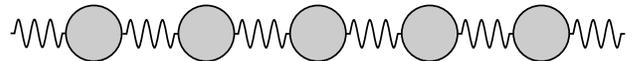}
\caption{A one-dimensional harmonic crystal}
\label{FAST-fig-crystal5}
\end{figure}

Consider the following system of ordinary differential equations:
%\cite{kloeden1999,stepanov2013stochastic}:
\begin{gather}
\dt v_i = F_i,  %+ b_i d W_i,
 \qquad
\dt u_i = v_i,
\label{FAST-1}
\end{gather}
where
\begin{gather}
F_i=\LL_{i} u_i - \eta v_i,
\label{FAST-F_i}
\\
%d W_i= \rho_i \sqrt{dt}
%\label{FAST-Winer},\\
\w_0 \= \sqrt{C/m}.
\end{gather}
Here $i$ is an
arbitrary integer which
describes the position of a particle in the chain; 
%the stochastic processes 
$u_i(t)$ and $v_i(t)$ are the displacement and
the particle velocity, respectively;
$F_i$ is the specific force on the
particle;  
%$W_i$ are Wiener processes;
%$b_i(t)$ is the intensity of the random external excitation;
$\eta$ is the specific viscosity for the environment;
$C$ is the bond stiffness; $m$ is the mass of a particle;
$\dt$ is the operator of differentiation with respect to time;
{$\L_i$} is the linear finite
difference operator:
\begin{gather}
\L_i u_i=u_{i+1} - 2u_{i} + u_{i-1}.
\label{FAST-Li}
\end{gather}
%Note that the results of the paper can be generalized for more
%complex finite difference operators and related physical systems (e.g. a
%crystal on an elastic support, next neighbour interactions etc).}
System of ODE 
\eqref{FAST-1} describes the motions of 
one-dimensional harmonic crystal (an ordered chain of identical interacting
material particles, see Fig.~\ref{FAST-fig-crystal5}).

%The normal random variables $\rho_i$ are such that 
%\begin{equation}
%    \av{\rho_i} = 0, \qquad
%    \av{\rho_i\rho_j} = \de_{ij},
%    \label{FAST-82}
%\end{equation}
%and they are assumed to be independent of $u_i$ and $v_i$.
The initial conditions are as follows: for all $i$,
\begin{gather}
u_i(0) =0,\qquad 
v_i(0)=\rho_i,
\label{FAST-ic-stochastic-kao1}
\end{gather}
where
the normal random variables $\rho_i$ are such that 
\begin{gather}
\langle \rho_i(0)\rangle=0,
%\langle u_i(0)u_j(0)\rangle =0,\\
\qquad
\langle \rho_i(0)\rho_j(0)\rangle=\kao\delta_{ij},
\label{FAST-ic-stochastic-kao2}
%\\
%\langle u_i(0)v_j(0)\rangle =0
%\label{FAST-ic-stochastic-mean}
\end{gather}
where $\kappa_{i}(0)$ is a given function of $i\in\mathbb Z$. {Later,
in Section~\ref{FAST-sec-uniform}, it will be assumed that $\kappa_{i}(0)$ do
not depend on $i$. It is useful to proceed with the derivation of basic
equations in Section~\ref{FAST-SSec-covariance}
not taking into account this supposition.} 
Note that in the latter more general case, the boundary conditions at the
infinity may be needed. These boundary conditions should guarantee that there
are no sources at the infinity, or give a mathematical description for such a
source. In the case under consideration in the paper, these special boundary
conditions are not necessary, we will use the requirement of spatial uniformity for
all physical quantities instead.  At the same time, in numerical calculations
(see Section~\ref{FAST-sec-numerics}), where we deal with a model for a finite harmonic
crystal, we use periodic boundary conditions 
\eqref{FAST-bc-periodic}
to provide the spatial
uniformity.

%In the case $b_i\equiv b$, equations \eqref{FAST-1} are the Langevin equations
%\cite{langevin1908theorie,lemons1997paul} for a
%one-dimensional harmonic crystal (an ordered chain of identical interacting
%material particles, see Fig.~\ref{fig-crystal})
%surrounded by a viscous environment (e.g., a gas or a liquid). Assuming that
%$b_i$ may depend on $i$, we introduce a natural generalization of the Langevin
%equation which allows one to describe the possibility of an external heat
%excitation (e.g.,\ laser excitation). 
%\NEW{This external excitation is assumed to be localized in space (in
%the paper we mostly consider the case of a point heat source) 
%and much more intensive than the stochastic influence caused by a non-zero
%temperature of the environment. Therefore,
%we neglect 
%in 
%\eqref{FAST-1}
%the constant stochastic term that does not depend on $i$}.
%Note that since $b_i$ do not depend on
%$u_j$ and $v_j$ for all $i,j$, it is not necessary to distinguish between the Stratonovich and It\^o
%formalism \cite{kloeden1999} in the case of equation \eqref{FAST-1}.

\subsection{The dynamics of covariances}
\label{FAST-SSec-covariance}

According to \eqref{FAST-F_i}, $F_i$ are linear functions of $u_i,\ v_i$. Taking
this fact into account together with Eqs.~\eqref{FAST-ic-stochastic-kao1},
\eqref{FAST-ic-stochastic-kao2},
we see that for all $t$
\begin{equation}
\av{u_i}=0,\qquad \av{v_i}=0.
\end{equation}
Following \cite{krivtsov2016ioffe}, consider the infinite sets of
covariance variables
\begin{equation}
    \xi_{p,q} \= \av{u_p u_q}, \qquad
    \nu_{p,q} \= \av{u_p v_q}, \qquad
    \ka_{p,q} \= \av{v_p v_q}.
\label{FAST-4}
\end{equation}
%and the \NEW{quantities}
%\begin{equation}
%\hat\beta_{p,q} \= \de_{pq}\,b_p b_q.
%\label{FAST-beta-pq}
%\end{equation}
%In the last equation, we take into account the second equation of \eqref{FAST-82}.
Thus, the variables $\xi_{p,q},\ \nu_{p,q},\ \ka_{p,q}$ 
%and $\hat\beta_{p,q}$}
are defined
for any pair of crystal particles.

We call the quantity
\begin{equation}
\mathscr K_{p,q}=\frac{m\kappa_{p,q}}2
\label{FAST-genkin}
\end{equation}
the generalized kinetic energy.
It is clear that quantities $m\kappa_{p,p}$ equal the estimated value for
doubled kinetic energy $2\mathscr K_{p,p}$ for particle with number $p$.
Accordingly, we identify the following quantities
\begin{gather}
     T_p \= 2k_B^{-1}\mathscr K_{p,p}= mk_B^{-1} \ka_{p,p}
%%% ,
\label{FAST-temp-def}     
%%% \\
%%%      \chi_i \= \tfrac12m k_B^{-1} \beta_{i,i}.
%%% \label{FAST-chi-def}
\end{gather}
as the kinetic temperature.
Here $k_B$ is the Boltzmann constant. 

For simplicity, in what follows, we drop the
subscripts $p$ and $q$, i.e., $\xi\=\xi_{p,q}$ etc. By definition, we also put
$\xi^\top\=\xi_{q,p}$ etc. 
Now we differentiate variables 
\eqref{FAST-4} with respect to time
taking into account equations of motion \eqref{FAST-1}. This yields
the following closed system of differential equations for
covariances:
\begin{gather}	
    \dt\xi = \nu + \nu^\top,
\label{FAST-5-1}
\\
    \dt\nu + \eta\nu = \LL_q\xi + \ka,
\label{FAST-5-2}
    \\
    \dt\ka + 2\eta\ka = \LL_p\nu + \LL_q\nu^\top,
\label{FAST-5-3}
\end{gather}
%\label{FAST-5}
where {$\L_p$ and $\L_q$ are the linear difference operators defined by 
Eq.~\eqref{FAST-Li} that act on  
$\xi_{p,q},\ \nu_{p,q},\ \ka_{p,q},\ \beta_{p,q}$
with respect to the first index subscript $p$ and the second one $q$, respectively.}
The initial conditions that correspond to 
Eqs.~\eqref{FAST-ic-stochastic-kao1}, \eqref{FAST-ic-stochastic-kao2} are
\begin{gather}
\xi_{pq}(0)=0,
\qquad
\nu_{pq}(0)=0,
\qquad
\kappa_{pq}(0)=\kappa_{p}(0)\delta_{pq}.
\end{gather}
Taking into account these initial conditions, it is useful to rewrite 
Eq.~\eqref{FAST-5-3} in the following form
\begin{equation}
    \dt\ka + 2\eta\ka = \LL_p\nu + \LL_q\nu^\top + \beta,
\label{FAST-5-3-mod}
\end{equation}
where singular term $\beta$ is
\begin{equation}
%\beta=\hat\beta+\kappa_{pp}(0)\delta_{pq}\delta(t),
\beta=\kappa_{p}(0)\delta_{pq}\delta(t).
\label{FAST-beta-delta}
\end{equation}
Equations~\eqref{FAST-5-1}, \eqref{FAST-5-2}, \eqref{FAST-5-3-mod} should be
supplemented 
with initial conditions in the following form, 
which is conventional for distributions (or generalized functions)~\cite{Vladimirov1971}:
\begin{equation}
\xi\big|_{t<0}\equiv0,
\qquad
\nu\big|_{t<0}\equiv0,
\qquad
\kappa\big|_{t<0}\equiv0.
\label{FAST-ic<0-pre}
\end{equation}

Now we introduce the symmetric and antisymmetric difference operators
\begin{equation}
2\L^{\mathrm S} \= \L_p+\L_q,\qquad
2\L^{\mathrm A} \= \L_p-\L_q,
\end{equation}
and the symmetric and antisymmetric parts of the variable $\nu$:
\begin{equation}
    2\nu^{\mathrm S} \= \nu+\nu^\top, \qquad
    2\nu^{\mathrm A} \= \nu-\nu^\top.
\label{FAST-9}
\end{equation}
Note that $\xi$ and $\kappa$ are symmetric variables.
Now Eqs.~\eqref{FAST-5-1}, \eqref{FAST-5-1}, \eqref{FAST-5-3-mod} can be rewritten as follows: 
\begin{gather}
\dt\xi = 2\nu^{\mathrm S},
\qquad(\dt + 2\eta)\ka =  2\LL^{\mathrm S}\nu^{\mathrm S} 
+ 2\LL^{\mathrm A}\nu^{\mathrm A} + \beta,
\label{FAST-11}
\\
(\dt + \eta)\nu^{\mathrm A} = -\LL^{\mathrm A}\xi,
\qquad(\dt + \eta)\nu^{\mathrm S} = \LL^{\mathrm S}\xi + \ka.
\label{FAST-11-post}
\end{gather}
This system of equations can be reduced 
(see \cite{gavrilov2018heat})
to one equation of the fourth order in
time for covariances of the particle velocities $\kappa$
\iffalse
%%% Applying the operator $\dt + \eta$ to Eqs.~\eqref{FAST-11} and
%%% substituting expressions
%%% \eqref{FAST-11-post}
%%% yields a closed system
%%% of two equations of second order in time:
%%% \begin{gather}	
%%%     \dt(\dt + \eta)\xi = 2(\LL^{\mathrm S}\xi + \ka),\label{FAST-12--1}\\
%%%     (\dt + \eta)(\dt + 2\eta)\ka 
%%%     = 2\big((\LL^{\mathrm S})^2-(\LL^{\mathrm  A})^2\big)\xi 
%%%     + 2\LL^{\mathrm S}\ka + (\dt + \eta)\beta.
%%% \label{FAST-12--2}
%%% \end{gather}
%%% \NEW{We can express $\ka$ in terms of $\xi$ using Eq.~\eqref{FAST-12--1}:
%%% \begin{equation}
%%%     \ka  = \frac12(\dt^2 + \eta\dt - 2\LL^{\mathrm S})\xi,
%%% \label{FAST-29}
%%% \end{equation}
%%% and substitute the result into Eq.~\eqref{FAST-12--2}. This yields 
%%% \begin{multline}
%%%     \frac12\Big(\big((\dt + \eta)(\dt + 2\eta)-
%%%      2\LL^{\mathrm S}\big)(\dt^2 + \eta\dt - 2\LL^{\mathrm S})
%%%      - 4\big((\LL^{\mathrm S})^2-(\LL^{\mathrm  A})^2\Big)\xi 
%%%     \\= (\dt + \eta)\beta.
%%% \label{FAST-podrob}
%%% \end{multline}
%%% Simplifying the left-hand side of 
%%% Eq.~\eqref{FAST-podrob}
%%% results in
%%% an equation of
%%% fourth order in time for $\xi$:
%%% Now we apply the operator 
%%% $\frac12(\dt^2 + \eta\dt - 2\LL^{\mathrm S})$
%%% to Eq.~\eqref{FAST-13}. Taking into account~\eqref{FAST-29},
%%% this yields a fourth-order equation for the covariances of the particle
%%% velocities~$\ka$:}%
\fi
\begin{multline}
\((\dt + \eta)^2(\dt^2 + 2\eta\dt - 4
\LL^{\mathrm S}) 
+ 4
(\LL^{\mathrm A})^2\)\ka 
\\=
(\dt+\eta)(\dt^2 + \eta\dt - 2
\LL^{\mathrm S})\beta,
\label{FAST-30}
\end{multline}
or, alternatively,
to one equation of the fourth order in
time for covariances of the displacements $\xi$:
\begin{equation}
\big((\dt + \eta)^2(\dt^2 + 2\eta\dt - 4\LL^{\mathrm S}) 
+ 4(\LL^{\mathrm A})^2\big)\xi =
    2(\dt + \eta)\beta.
\label{FAST-13}
\end{equation}

%In what follows, we deal with Eq.~\eqref{FAST-30}. 
%To take into account non-zero classical initial conditions, one needs to add 
%the corresponding singular terms (in the form of a linear combination of 
%$\delta(t)$ and its derivatives) to the right-hand sides of the corresponding equations 
%\cite{Vladimirov1971}.

%\NEW{Let us note that  equation \eqref{FAST-30} is
%a determenistic equation}. What is also important is that \eqref{FAST-30} is a closed
%equation. Thus the thermal processes do not depend on any property of the
%cumulative distribution functions for the displacements and the particle
%velocities other than the covariance variables used above.

\section{The case of a uniform initial kinetic temperature distribution}
\label{FAST-sec-uniform}

Following \cite{krivtsov2015heat,krivtsov2016ioffe}, we 
introduce the discrete spatial variable
\begin{equation}
k\= p+q
\label{FAST-k-def}
\end{equation}
and the discrete correlational variable
%\TODO{уже введен после \eqref{f12}}
\begin{equation}
     n\=q-p
\label{FAST-n}
\end{equation}
instead of discrete variables $p$ and $q$.
We have
\begin{equation}
q=\frac k2+\frac n2,\qquad p=\frac k2-\frac n2.
\label{FAST-pq-nk}
\end{equation}

In what follows, we consider the case when the initial values
$\kappa_{p}(0)\delta_{pq}$ of
the covariance variables $\kappa_{pq}$ do not depend on the spatial variable
$k$, and depend only
on $n$, i.e.\
\begin{equation}
\kappa_{p}(0)=\bar\kappa_0,
\label{FAST-UNIFORM}
\end{equation}
where $\bar\kappa_0$ is a given constant. From the physical point of view this
means that we have a uniform distribution of the initial value for the
kinetic temperature 
\begin{equation}
T_p\big|_{t=0}=\bar T_0\equiv2 k_B^{-1}\mkE,
\end{equation}
where 
\begin{equation}
\mkE\equiv \frac{m\bar\kappa_0}2
\label{FAST-mkE-def}
\end{equation}
is the initial value for both the total and the kinetic energy.
%\TODO{How can this be realized?}
%Due to symmetry, 
In the case of a uniform initial conditions 
it is natural to assume that for $t>0$ we also have
\begin{equation}
\kappa_{pq}=\hat\kappa_n,\qquad \xi_{pq}=\hat\xi_{n}.
\label{FAST-all-n}
\end{equation}

In the case 
\eqref{FAST-UNIFORM}
of the uniform initial kinetic temperature distribution, 
in order to measure the estimated value of the (doubled) potential energy, we introduce the
following quantity
\begin{equation}	
4\Pi_{p,p}=
m\omega_0^2\left(
\left\langle(u_{p-1}-u_p)^2\right\rangle
+
\left\langle(u_{p+1}-u_p)^2\right\rangle
\right).
\label{FAST-potential-def}
\end{equation}
Provided that 
\eqref{FAST-all-n} are true, one has 
\begin{multline}	
4\Pi_{p,p}\equiv2\hat\Pi_0\\=
m\omega_0^2\big(2\xi_{p,p}
+\xi_{p-1,p-1}
-2\xi_{p,p-1}
+\xi_{p+1,p+1}
-2\xi_{p,p+1}\big)\\=
-2m\omega_0^2\L_n\hat\xi_n\big|_{n=0}.
\label{FAST-Pi-calc-formula}
\end{multline}
% % %According to
% % %Eq.~\eqref{FAST-potential-def}
We call quantities $\Pi_{p,q}\equiv\hat\Pi_n=-\frac12m\omega_0^2\L_n\hat\xi_n$
the generalized potential energy.
The following identities are true
for any quantity $\zeta_{p,q}$ such that 
$
\zeta_{p,q}=\zeta_{q,p}=\hat\zeta_n:
$
\begin{gather}
\begin{multlined}	
2\L^{\mathrm S}\zeta_{p,q}=
\zeta_{p+1,q}
+\zeta_{p-1,q}
-4\zeta_{p,q}
+\zeta_{p,q+1}
+\zeta_{p,q-1}
\\=
2\zeta_{p+1,q}
-4\zeta_{p,q}
+2\zeta_{p-1,q}=2\L_n\hat\zeta_n,
\end{multlined}
\\
2\L^{\mathrm A}\zeta_{p,q}=
\zeta_{p+1,q}
+\zeta_{p-1,q}
-\zeta_{p,q+1}
-\zeta_{p,q-1}=0.
\end{gather}
Now, taking into account 
Eqs.~\eqref{FAST-ic<0-pre},
\eqref{FAST-beta-delta},
Eqs.~\eqref{FAST-30}, \eqref{FAST-13},
can be rewritten as
\begin{gather}
\begin{multlined}	
(\dt + \eta)(\dt^2 + 2\eta\dt - 4
\LL_n) 
%%% + 4
%%% (\LL^{\mathrm A})^2
\hat{\mathscr K}_n 
\\\qquad\qquad=
{\mkE}
(\dt^2 + \eta\dt - 2
\LL_n)\delta(t)\delta_n,
\end{multlined}
\label{FAST-30-mod}\\
(\dt + \eta)(\dt^2 + 2\eta\dt - 4\LL_n) 
%+ 4(\LL^{\mathrm A})^2
%\big
\hat\Pi_n 
=-2{\mkE}\omega_0^2
%(\dt + \eta)
\delta(t)\L_n\delta_n,
\label{FAST-13-mod}
\end{gather}
respectively.

It is useful to consider also the following quantities
\begin{equation}
\hat\la_n\equiv\hat{\mathscr K}_n-\hat\Pi_n
\label{FAST-lagrangian-def}
\end{equation}
and
\begin{equation}
\hat{\mathscr E}_n\equiv
\hat{\mathscr K}_n+\hat\Pi_n.
\label{FAST-E-def}
\end{equation}
We call these quantities the generalized Lagrangian and the generalized total
energy, respectively.
%,
%since $\hat\Lambda_0$ and 
%$\hat{\mathscr E}_0$ are the Lagrangian and the total energy for an elementary
%element of the harmonic crystal. 
The corresponding equations for these quantities can be obtained by means of applying of the operator $\L_n$ to 
\eqref{FAST-13-mod} and calculating the sum or difference of both parts of
Eqs.~\eqref{FAST-30-mod}, \eqref{FAST-13-mod}. 
%This yields 
%\begin{gather}
%    (\dt + \eta)(\dt^2 + 2\eta\dt - 4\LL_n)\hat\la_n =m\kappa_0
%    \dt(\dt+\eta)\delta(t)\delta_n,
%\label{FAST-f4}\\
%    (\dt + \eta)(\dt^2 + 2\eta\dt - 4\LL_n)\hat{\mathscr E}_n =m\kappa_0
%(\dt^2 + \eta\dt - 4\LL_n)
%    \delta(t)\delta_n,
%\end{gather}
%respectively. 
Taking into account 
initial condition {in the form of}
\eqref{FAST-ic<0-pre}, this yields
\begin{gather}
    (\dt^2 + 2\eta\dt - 4\LL_n)\hat\la_n =
    {\mkE}\dt \delta(t)\delta_n,
\label{FAST-f4}\\
\begin{multlined}	
    (\dt + \eta)(\dt^2 + 2\eta\dt - 4\LL_n)
    \hat{\mathscr E}_n 
\\
\qquad\qquad=
    {\mkE}(\dt^2 + \eta\dt - 4\LL_n)\delta(t)\delta_n, 
\end{multlined}
\label{FAST-En}
\end{gather}
respectively. The corresponding initial conditions are
\begin{equation}
\hat\Lambda_n\big|_{t<0}\equiv0,
\qquad
\hat{\mathscr E}_n\big|_{t<0}\equiv0.
\label{FAST-ic<0-LE}
\end{equation}

To calculate every energetic quantity from the set $\hat{\mathscr K}_0$, $\hat\Pi_0$,
$\hat \Lambda_0$, $\hat{\mathscr E}_0$ it is enough to solve any two equations
from the set 
\eqref{FAST-30-mod},
\eqref{FAST-13-mod},
\eqref{FAST-f4},
\eqref{FAST-En}. In what follows, we deal with 
Eqs.~\eqref{FAST-f4},
\eqref{FAST-13-mod} which have a simpler structure. The generalized kinetic
energy and the generalized total energy in this case can be calculated as
follows:
\begin{gather}
\hat{\mathscr K}_n=\hat\Lambda_n
+\hat\Pi_n,
\label{FAST-K-calc}
\\
\hat{\mathscr E}_n=\hat\Lambda_n
+2\hat\Pi_n.
\label{FAST-E-calc}
\end{gather}

%%% Considering the fast motions it is useful to consider instead of 
%%% equation \eqref{30}
%%% the equation for
%%% the generalized Lagrangian
%%% %\TODO{\foreignlanguage{russian}{А если все-таки для $\kappa$?}}
%%% %\TODO{\foreignlanguage{russian}{Почему это лагранжиан?}}
%%% Due to \eqref{13}, \eqref{30}, 
%%% \eqref{chi-def},
%%% \eqref{A2-zero}
%%% the equation of the fourth order for $\la_n$ is
%\TODO{\foreignlanguage{russian}{Вопрос о двойке: результаты мои и Антона,
%по-видимому идентичны, т.к. у него в НУ в статье ДАН стоит полная энергия,
%равная удвоенной кинетической.}}
%%% We consider this equation supplemented with zero initial conditions formulated in the following form
%%% \begin{equation}
%%% \la_n(t)\big|_{t<0}\equiv0.
%%% \label{ic-fast}
%%% \end{equation}
%%% Integrating \eqref{f4}
%%% taking into account 
%%% \eqref{ic-fast}
%%% yields the  following infinite system of ordinary differential equations
%%% \begin{equation}
%%%     \ddot\la_n + 2\eta\dot\la_n - 4\LL_n\la_n =
%%%     2\dot\chi\delta_n.
%%% \label{maineq-fast}
%%% \end{equation}

%где константы интегрирования \DANGER{отброшены}\footnote{\it Обосновать!}.
% В случае постоянного $\beta$ получаем
% \ba{f6}{c}
%     \ddot\la + 2\eta\dot\la - 4\L\la = 0.
% \ea

\section{Solution of the equations for energetic quantities}% for fast motions}
\label{FAST-Sec-analyt-intform}
\let\La=\varLambda

%In what follows, we investigate the initial value problem for the system of
%ordinary differential equations \eqref{maineq-fast},
%wherein 
%\begin{equation}
%\chi=\delta(t), 
%\end{equation}
%with zero initial conditions formulated in the following form
%\begin{equation}
%\Lambda_n(t)\big|_{t<0}\equiv0.
%\label{ic<0-fast}
%\end{equation}
%This corresponds to choice of the heat supply in the form of a point
%pulse source with ordinary in space intensity. 

%\TODO{\foreignlanguage{russian}{$\omega_0$ --- большой параметр, обсудить}}

\subsection{The Lagrangian $\hat\Lambda_0$}
%Take $\eta=O(1)>0$.  Taking $\Psi=\Lambda_n$ in 
%\aeqref{0-rule}, one can easily show that 

%In this section and in what follows we use more compact notation: the overdot means
%$\partial_t$. 
We apply the discrete-time Fourier transform 
%$\mathscr F_n^q$ 
\cite{proakis1996digital,Slepian1980} with respect to the variable $n$
to 
%$\theta_n(x,t)$,
Eq.~\eqref{FAST-f4}. 
This yields
\begin{gather}
\dt^2\F{\Lambda} q n
+2\eta\dt\F{\Lambda} q n
% -
+ %\Big(16\omega_0^2\sin^2\frac q2\Big)
\mathscr A^2\F\Lambda q n
={\mkE}\dt\delta(t),
\label{FAST-maineq-fast-trans-eta}
\\
\mathscr A=4\omega_0\Big|\sin\frac \q2\Big|,
\label{FAST-A-def}
\end{gather}
where
\begin{gather}
\F\Lambda{\q}n(\q,t)=\sum_{n=-\infty}^{\infty} \hat\Lambda_n \exp (-\I n\q).
\end{gather}
Here and in what follows, $\q$ is the wavenumber, $\I$ is the imaginary unit.
In order to obtain Eqs.~\eqref{FAST-maineq-fast-trans-eta},
\eqref{FAST-A-def}
we used the shift property \cite{proakis1996digital} of the discrete-time Fourier transform:
\begin{equation}
%\F\Lambda{q}n(q,t)
\sum_{n=-\infty}^{\infty} \hat\Lambda_{n\pm1} \exp (-\I n\q)
=\exp(\pm \I \q)\,\F\Lambda \q n(\q,t).
%\sum_{n=-\infty}^{\infty} \hat\Lambda_n \exp (-inq).
\label{FAST-shift}
\end{equation}

Equation 
\eqref{FAST-maineq-fast-trans-eta}
together with initial conditions in the form of Eq.~\eqref{FAST-ic<0-LE} is 
equivalent \cite{Vladimirov1971} to initial value problem
for the corresponding homogeneous equation 
with the following classical initial conditions:
\begin{equation}
\begin{gathered}
\F\Lambda{}{}\big|_{t=0}={\mkE},\\	
\dt\F{\Lambda}{}{}\big|_{t=0}=-2\eta {\mkE}.	
\end{gathered}
\end{equation}
The corresponding solution is
\begin{equation}
\begin{gathered}
\F{\Lambda}{\q}n(\q,t)=
\left\{
\begin{aligned}	
&
%2
%\Lambda_F^{(0)}
%+
%2
\Lambda_F^{(2)}
%\cos \big( \sqrt {\mathscr A^2-{{\eta}}^{2}}\,t\big)
,
&{\bar \q}<|\q|\leq\pi
;\\
&
%2
%\Lambda_F^{(0)}
%+
%2
\Lambda_F^{(1)}
,
&|\q|<{\bar \q}
,
\end{aligned}
\right.
,
\\
\begin{multlined}	
\F\Lambda{\q}n^{(2)}(\q,t)={\mkE}{{\mathrm e}^{-{\eta}\,t}}
\left(
-\frac {{\eta}\,
\sin  \big( \sqrt {\mathscr A^2-\eta^2}\,t \big) }
        {\sqrt {\mathscr A^2-\eta^2}}
\right.
\\+
\left.
\cos \big( \sqrt {\mathscr A^2-{{\eta}}^{2}}\,t
 \big)
\right)
,\end{multlined}
\\
\begin{multlined}	
\F\Lambda{\q}n^{(1)}(\q,t)={\mkE}{{\mathrm e}^{-{\eta}\,t}}
\\\times 
\left(
\sum_{(\pm)}
\frac
{ 
\big( {\mathscr A}^{2}\pm\eta\sqrt {{\eta}^{2}-{\mathscr A}^{2}}-{\eta}^{2} \big) 
{\mathrm e}^{\big(\pm\sqrt {{\eta}^{2}-{\mathscr A}^{2}}\big) t}
}
{2({\mathscr A}^{2}-{{\eta}}^{2})}
\right)
.
\end{multlined}
\end{gathered}
\label{FAST-lambdaF-e}
\end{equation}
%\begin{equation}
%\F\Lambda{q}n(q,t)={\mkE}{{\mathrm e}^{-{\eta}\,t}}
%\left\{
%\begin{aligned}	
%&-\frac {{\eta}\,
%\sin  \big( \sqrt {\mathscr A^2-\eta^2}\,t \big) }
%        {\sqrt {\mathscr A^2-\eta^2}}
%+\cos \big( \sqrt {\mathscr A^2-{{\eta}}^{2}}\,t
% \big),
%&{\bar q}<|q|\leq\pi
%;\\
%&
%\sum_{(\pm)}
%\frac
%{ 
%\big( {\mathscr A}^{2}\pm\eta\sqrt {{\eta}^{2}-{\mathscr A}^{2}}-{\eta}^{2} \big) 
%{\mathrm e}^{\big(\pm\sqrt {{\eta}^{2}-{\mathscr A}^{2}}\big) t}
%}
%{2({\mathscr A}^{2}-{{\eta}}^{2})}
%%+{\frac { \left( {\mathscr A}^{2}-\sqrt {-{\mathscr A}^{2}+{\eta}^{
%%2}}\eta-{\eta}^{2} \right) {{e}^{- \left( \eta+
%%\sqrt {-{\mathscr A}^{2}} \right) t}}}{{\mathscr A}^{2}-{\eta}^{2}}}
%,
%&|q|<{\bar q}
%,
%\end{aligned}
%\right.
%\label{FAST-lambdaF-e}
%\end{equation}
where
\begin{equation}
{\bar \q}=
2\arcsin{\frac\eta{4\omega_0}}.
\end{equation}
In what follows, we generally assume that the underdamped case 
%$\eta/\omega_0\ll1$ \cite{gavrilov2018heat},
%therefore we assume}
\begin{equation}
\eta<4\omega_0 \quad\Longleftrightarrow\quad {\bar \q}<\pi
\label{FAST-eta-restriction}
\end{equation}
is under consideration. The critically damped and the overdamped cases are briefly discussed in 
Section~\ref{FAST-Sec-overdamped} (see also Figures~\ref{fast_critical.eps},
\ref{fast_overdamped.eps}).
Now we apply the inverse transform
\begin{equation}
\hat\Lambda_n=\frac1{2\pi}\int_{-\pi}^{\pi}\F\Lambda{\q}n\,\exp(\I n\q)\,\d \q
%=
%\frac1{2\pi}
%\left(
%\int_{-\bar q}^{\bar q}
%+\int_{-\pi}^{-\bar q}
%+\int_{\bar q}^\pi
%\right)
%\F\Lambda{q}n\,\cos(n\q)\,\d q
%\\=
%\frac1{2\pi}
%\left(
%\int_{-\bar q}^{\bar q}
%+
%\int_{\pi-\bar q}^{\pi+\bar q}
%\right)
%\F\Lambda{q}n\,\cos(n\q)\,\d q
%\equiv
%\hat\Lambda_n^{(1)}
%+
%\hat\Lambda_n^{(2)}
\label{FAST-DTFT-inversion}
\end{equation}
and get the solution in the integral form for $\hat\Lambda_n$. The
Lagrangian $\hat\Lambda_0$ equals
\begin{equation}
\hat\Lambda_0=\frac1{2\pi}\int_{-\pi}^{\pi}\F\Lambda{\q}n\,\d \q.
\label{FAST-DTFT-inversion-l0}
\end{equation}

\subsection{The potential energy $\hat \Pi_0$}
Applying the discrete-time Fourier transform 
%$\mathscr F_n^q$ 
%\cite{Slepian1980} with respect to the variable $n$
to 
%$\theta_n(x,t)$,
Eq.~\eqref{FAST-13-mod}
yields
\begin{equation}
\dt^3\Pi_F+2\eta\dt^2\Pi_F+\mathscr A^2\dt\Pi_F=8{\mkE}\omega_0^2
\sin^2\frac \q2\,\delta(t).
\label{FAST-dddot}
\end{equation}
Equation 
\eqref{FAST-dddot}
together with initial conditions in the form of Eq.~\eqref{FAST-ic<0-LE} is 
equivalent \cite{Vladimirov1971} to the initial value problem
for the corresponding homogeneous equation 
with the following classical initial conditions:
\begin{equation}
\begin{gathered}
\F\Pi{}{}\big|_{t=0}=0,\\	
\dt\F{\Pi}{}{}\big|_{t=0}=0,\\
\dt^2\F{\Pi}{}{}\big|_{t=0}=8{\mkE}\omega_0^2\sin^2\frac \q2.	
\end{gathered}
\end{equation}
The corresponding solution
%is
%\begin{equation}
%\F{\Pi}{q}n(q,t)=
%\frac{
%{\mkE} \mathscr A^2\,
%{{\mathrm e}^{-{\eta}\,t}}}
%{2({\mathscr A}^{2}-{{\eta}}^{2})}
%\left\{
%\begin{aligned}	
%&
%%\frac{
%%8m\omega_0^2 \kappa_0 \sin^2\frac \q2\,
%%{{\mathrm e}^{-{\eta}\,t}}
%%\Big(
%1-\cos \big( \sqrt {\mathscr A^2-{{\eta}}^{2}}\,t\big)
%%\Big)
%%}
%%{\mathscr A^2-\eta^2}
%,
%&{\bar q}<|q|\leq\pi
%;\\
%&
%%\frac
%%{8m\omega_0^2 \kappa_0 \sin^2\tfrac \q2\,
%%\mathrm e ^{-\eta t}}
%%{2({\mathscr A}^{2}-{{\eta}}^{2})}
%%\left(
%1-
%%\sum_{(\pm)}
%{ 
%%\big( {\mathscr A}^{2}\pm\eta\sqrt {{\eta}^{2}-{\mathscr A}^{2}}-{\eta}^{2} \big) 
%\cosh {\big(\sqrt {{\eta}^{2}-{\mathscr A}^{2}}\,t\big) }
%}
%%\right)
%%+{\frac { \left( {\mathscr A}^{2}-\sqrt {-{\mathscr A}^{2}+{\eta}^{
%%2}}\eta-{\eta}^{2} \right) {{e}^{- \left( \eta+
%%\sqrt {-{\mathscr A}^{2}} \right) t}}}{{\mathscr A}^{2}-{\eta}^{2}}}
%,
%&|q|<{\bar q}
%,
%\end{aligned}
%\right.
%\label{FAST-PiF-e}
%\end{equation}
%Expression
%\eqref{FAST-PiF-e}
can be written as follows:
\begin{gather}	
\F{\Pi}{\q}n(\q,t)=
\left\{
\begin{aligned}	
&
%2
\Pi_F^{(0)}
+
%2
\Pi_F^{(2)}
%\cos \big( \sqrt {\mathscr A^2-{{\eta}}^{2}}\,t\big)
,
&{\bar \q}<|\q|\leq\pi
;
\label{FAST-PiF-e}
\\
&
%2
\Pi_F^{(0)}
+
%2
\Pi_F^{(1)}
,
&|\q|<{\bar \q}
,
\end{aligned}
\right.
,
\\
\Pi_F^{(0)}=
\frac{
{\mkE} \mathscr A^2\,
{{\mathrm e}^{-{\eta}\,t}}}
{2({\mathscr A}^{2}-{{\eta}}^{2})}
\label{FAST-def-Pi0F}
,\\
\Pi_F^{(1)}=-
\frac{
{\mkE} \mathscr A^2\,
{{\mathrm e}^{-{\eta}\,t}}}
{2({\mathscr A}^{2}-{{\eta}}^{2})}
%\frac12
%\sum_{(\pm)}
{ 
{\cosh\big(\sqrt {{\eta}^{2}-{\mathscr A}^{2}}\,t\big) }
}
,\\
\Pi_F^{(2)}=-
\frac{
{\mkE} \mathscr A^2\,
{{\mathrm e}^{-{\eta}\,t}}}
{2({\mathscr A}^{2}-{{\eta}}^{2})}
\cos \big( \sqrt {\mathscr A^2-{{\eta}}^{2}}\,t\big).
\end{gather}

Now we apply the inverse transform
\begin{equation}
\hat\Pi_n=\frac1{2\pi}\int_{-\pi}^{\pi}\F\Pi{q}n\,\exp(\I n\q)\,\d \q
\label{FAST-DTFT-inversion-Pi}
\end{equation}
and get the solution in the integral form for $\hat\Pi_n$. The
potential energy $\hat\Pi_0$ equals
\begin{equation}
\hat\Pi_0=\frac1{2\pi}\int_{-\pi}^{\pi}\F\Pi{\q}n\,\d \q.
\label{FAST-DTFT-inversion-Pi-l0}
\end{equation}

\subsection{The conservative case $\eta=0$}
\label{FAST-Sec-conservative}
%In this section we assume that $\eta=0$. 
%For all twice differentiable functions $\Psi(t)$ the following formulas are
%valid:
%\begin{equation}
%\begin{gathered}
%\partial_t \big(\Psi(t)H(t)\big)=\Psi(0)\delta(t)+\dot \Psi(t)H(t),\\
%\partial_t^2 \big(\Psi(t)H(t)\big)=\Psi(0)\dot\delta(t)+\dot\Psi(0)\delta(t)+\ddot \Psi(t)H(t).
%\end{gathered}
%\label{0-rule}
%\end{equation}
%Taking here $\Psi=\Lambda_n$, one can easily show that system of equations 
%\eqref{maineq-fast}
%together with initial conditions in form of Eq.~\eqref{ic<0-fast} is 
%equivalent \cite{Vladimirov1971} to initial value problem
%for the homogeneous equation corresponding to Eq.~\eqref{maineq-fast} 
%with following initial conditions:
%\begin{gather}
%\Lambda_n\big|_{t=0}=\delta_n,\\	
%\dot\Lambda_n\big|_{t=0}=0.	
%\end{gather}
%We apply the discrete-time Fourier transform $\mathscr F_n^q$ 
%\cite{Slepian1980} over the variable $n$
%to 
%%$\theta_n(x,t)$,
%Eq.~\eqref{maineq-fast}. 
%This yields
%\begin{gather}
%\F{\ddot\Lambda} q n
%%+2\eta\dot\theta_\F
%% -
%+ %\Big(16\omega_0^2\sin^2\frac \q2\Big)
%\mathscr A^2\F\Lambda q n
%=\dot\delta(t),
%\label{maineq-fast-trans}
%\end{gather}
%The solution of the ordinary differential equation 
%\eqref{maineq-fast-trans} that satisfy initial conditions in the form of 
%\eqref{ic<0-fast} is
%\begin{equation}
%\F\Lambda{q}n(q,t)=\cos \mathscr At
%%\Big(4\omega_0t\sin\frac \q2\Big).
%\end{equation}
%Now we apply the inverse transform \cite{?}
In the conservative case integral 
\eqref{FAST-DTFT-inversion} can be calculated in the closed form:
\begin{multline}
\hat\Lambda_n=\frac{{\mkE}}{\pi}\int_{0}^{\pi} \cos \mathscr At
%\Big(4\omega_0t\sin\frac \q2\Big)
\cos n\q\,\d \q\\=
\frac{2{\mkE}}{\pi}\int_{0}^{\pi/2} \cos(4\omega_0t\sin Q)\cos(2nQ)\,\d Q
\label{fast-calc-cons}
,
\end{multline}
This yields~\cite{PBM1}
\begin{equation}
\hat\Lambda_n=
{\mkE} J_{2n}(4\omega_0t).
\label{fast-solution0}
\end{equation}
For large time $t$ the asymptotics of the Bessel function $J_{2n}$ is well known
\cite{abramowitz1972handbook}, thus 
\eqref{fast-solution0} can be written in the asymptotic form
\begin{equation}
\hat\la_n = 
\frac{
{\mkE}
(-1)^n}{\sqrt{2\pi\w_0 t}}\,
\cos\Big(4\w_0 t-\frac\pi4\Big)+O\left(t^{-3/2}\right).
\label{f16}
\end{equation}

In the conservative case it is useful to take $\hat{\mathscr E}_n$ as the second
energetic variable (instead of $\hat\Pi_n$) since
Eq.~\eqref{FAST-En} simplifies to
\begin{gather}
    \dt
    \hat{\mathscr E}_n =
    {\mkE}\delta(t)\delta_n.
\label{FAST-En-0}
\end{gather}
The corresponding solution is
\begin{equation}
    \hat{\mathscr E}_n =
    {\mkE}\delta_n,
\end{equation}
i.e.\ the total energy $\hat{\mathscr E}_0$ conserves, keeping the initial
value for the
total (or kinetic) energy.

According to Eqs.~\eqref{FAST-lagrangian-def},
\eqref{FAST-E-def},
the generalized kinetic $\hat{\mathscr K}_n$ and potential 
$\hat{\Pi}_n$ energies equal
\begin{gather}
\begin{multlined}	
\hat{\mathscr K}_n=\frac{{\mkE}}2\big(\delta_n+J_{2n}(4\omega_0t)\big)
\\=\frac{{\mkE}}2\left(\delta_n+
\frac{(-1)^n}{\sqrt{2\pi\w_0 t}}\,
\cos\Big(4\w_0 t-\frac\pi4\Big)\right)+O\left(t^{-3/2}\right),
\end{multlined}
\\
\begin{multlined}	
\hat{\Pi}_n=\frac{{\mkE}}2\big(\delta_n-J_{2n}(4\omega_0t)\big)
\\=\frac{{\mkE}}2\left(\delta_n-
\frac{(-1)^n}{\sqrt{2\pi\w_0 t}}\,
\cos\Big(4\w_0 t-\frac\pi4\Big)\right)+O\left(t^{-3/2}\right)
.
\end{multlined}
\end{gather}
These results are in agreement with ones previously obtained in 
\cite{klein1953mecanique,krivtsov2014energy}.

\section{Asymptotics for the energetic quantities as $t\to\infty$}
\label{FAST-section-AS}
Unlike the conservative case the 
inverse Fourier transforms of quantities
\eqref{FAST-lambdaF-e} and
\eqref{FAST-PiF-e}
cannot be
evaluated in closed forms. 
Instead of this for large time we can proceed with
asymptotic estimation of the corresponding integrals. 

In this section we use the following notation
\begin{equation}
\omega\=\frac14{\sqrt{16\omega_0^2-\eta^2}}.
\end{equation}

\subsection{The Lagrangian $\hat\Lambda_0$}
Due to 
\eqref{FAST-lambdaF-e}
the integral in the right-hand side of 
\eqref{FAST-DTFT-inversion} can be represented as follows:
\begin{multline}
\hat\Lambda_n
%=\frac1{2\pi}\int_{-\pi}^{\pi}\F\Lambda{q}n\,\exp(\I n\q)\,\d q
=
\frac1{2\pi}
%\left(
\int_{-\bar \q}^{\bar \q}
\F\Lambda{\q}n\,\cos{n\q}\,\d \q
+
\frac1{\pi}
\int_{\bar \q}^\pi
%\right)
\F\Lambda{\q}n\,\cos{n\q}\,\d \q
%\\=
%\frac1{\pi}
%\left(
%\int_{0}^{\bar q}
%+
%\int_{\bar q}^{\pi}
%\right)
%\F\Lambda{q}n\,\cos{n\q}\,\d q
\\\equiv
\hat\Lambda_n^{(1)}
+
\hat\Lambda_n^{(2)}.
\label{FAST-DTFT-inversion-rep}
\end{multline}

It is easy to see that integral $\hat\Lambda_{n}^{(1)}$ defined by
Eqs.~\eqref{FAST-DTFT-inversion-rep}, \eqref{FAST-lambdaF-e}
is a sum of two Laplace type integrals and, therefore, it
can be estimated by the Laplace method
\cite{Fedoruk-Saddle}. 
On the other hand, integral $\hat\Lambda_{n}^{(2)}$ (also defined by
Eqs.~\eqref{FAST-DTFT-inversion-rep}, \eqref{FAST-lambdaF-e})
is a  Fourier type integral and therefore it can be estimated by 
the method of stationary phase \cite{Fedoruk-Saddle}. 

Calculation of the asymptotics for quantities $\hat\la_n^{(2)}$,
$\hat\la_n^{(1)}$ is
presented in Appendices~\ref{FAST-App-L2}--\ref{FAST-App-L1}, respectively. 
In the most interesting 
%from physical point of view 
particular case $n=0$,
{the result is} \footnote{In what follows, we will see
that to calculate the principal term of the asymptotics for 
$
\hat{\mathscr K}_n^{(1)}
$ we need to calculate two first non-zero terms of expansion for 
$
\hat{\la}_n^{(1)}
$
(see   
\eqref{FAST-as-K1} and text below the formula).}:
\begin{gather}
\begin{multlined}	
\hat\Lambda_0^{(1)}=
{\mkE}\left(
-
\frac{
t^{-3/2}}{8\sqrt{2\pi\eta}\,\omega_0}
%+
-
\frac{t^{-5/2}\big(
%(
%12n^2
%-
3
%)
\eta^2
%-
+
12\omega_0^2\big)\sqrt2}
{512\sqrt\pi\,\eta^{3/2}\omega_0^3}
\right)
\\+O(t^{-7/2}),
\end{multlined}
%+O\bigg(\frac {e^{-\eta t}}{t}\bigg).
\label{FAST-l2n}
%\\
%\hat\Lambda_n=\hat\Lambda_n^{(1)}+\hat\Lambda_n^{(2)}.
%\label{ln}
\\
\begin{multlined}
    \hat\la_0^{(2)} =
    %(-1)^n
    \frac{ {\mkE} \mathrm e^{-\eta t} }{2\omega_0\sqrt{2\pi t}}\,
    \Bigg(
2\sqrt\omega
    \cos\Big(4\omega t-\frac\pi4\Big)
    \\
    \qquad
    -\frac{\eta}{2\sqrt\omega}
    \sin\Big(4\omega t-\frac\pi4\Big)
    \Bigg)
+O\bigg(\frac {\mathrm e^{-\eta t}}{t}\bigg).
\label{FAST-as-final}
\end{multlined}
\end{gather}

\subsection{The potential energy $\hat \Pi_0$}

Calculation of the asymptotics for generalized potential energy $\hat\Pi_n$
is
presented in Appendices~\ref{FAST-App-P}--\ref{FAST-App-P0}, respectively. 
We have 
\begin{equation}
\hat\Pi_n
%=\frac1{2\pi}\int_{-\pi}^{\pi}\F\Lambda{q}n\,\exp(\I n\q)\,\d q
=
\hat\Pi_n^{(0)}
+
\hat\Pi_n^{(1)}
+
\hat\Pi_n^{(2)},
\label{FAST-DTFT-inversion-Pi-rep}
\end{equation}
where 
$\hat\Pi_n^{(1)}$ 
is a Laplace type integral, and
$\hat\Pi_n^{(2)}$
is a Fourier type integral.
For $n=0$ one gets
%\begin{multline}
%\hat\Pi_n
%=\frac1{2\pi}\int_{-\pi}^{\pi}\F\Lambda{q}n\,\exp(\I n\q)\,\d q
%=
%%%%\frac1{2\pi}
%%%%\left(
%%%%\int_{-\bar q}^{\bar q}
%%%%+\int_{-\pi}^{-\bar q}
%%%%+\int_{\bar q}^\pi
%%%%\right)
%%%%\F\Lambda{q}n\,\cos{n\q}\,\d q
%%%%\\=
%%%%\frac1{2\pi}
%%%%\left(
%%%%\underbrace{
%%%\frac1\pi
%%%\PV\int_{0}^{\pi}
%%%\Pi_F^{(0)}
%%%\cos{n\q}\,\d q
%%%%}_{\hat\Pi_n^{(0)}}
%%%\\+
%%%%\underbrace{
%%%\frac1\pi\PV\left(
%%%\int_{-\bar q}^{\bar q}
%%%\Pi_F^{(1)}
%%%\cos{n\q}\,\d q
%%%%}_{\hat\Pi_n^{(1)}}
%%%+
%%%%\underbrace{
%%%%\frac1\p
%%%\left(
%%%\int_{-\pi}^{-\bar q}
%%%+
%%%\int_{\bar q}^{\pi}
%%%\right)
%%%\Pi_F^{(2)}
%%%\cos{n\q}\,\d q
%%%%}_{\hat\Pi_n^{(2)}}
%%%\right)
%%%%%%\\+
%%%%%%2\lim_{\epsilon\to+0}
%%%%%%\left(
%%%%%%\int_{\bar q-\epsilon}^{\bar q}
%%%%%%\left(
%%%%%%\Pi_F^{(0)}
%%%%%%+
%%%%%%\Pi_F^{(1)}
%%%%%%\right)
%%%%%%\cos n\q
%%%%%%\,\d q
%%%%%%+
%%%%%%\int_{\bar q}^{\bar q+\epsilon}
%%%%%%\left(
%%%%%%\Pi_F^{(0)}
%%%%%%+
%%%%%%\Pi_F^{(1)}
%%%%%%\right)
%%%%%%\cos n\q
%%%%%%\,\d q
%%%%%%\right)
%%%%%%%\F\Lambda{q}n\,
%%%%%%%\cos{n\q}\,\d q
%%%\\
%%%\equiv
%+
%2\lim_{\epsilon\to+0}R_n(\epsilon,t),
%\label{FAST-DTFT-inversion-Pi-rep}
\begin{gather}
\hat\Pi_0^{(0)}
=
\frac1\pi
\PV\int_{0}^{\pi}
\Pi_F^{(0)}
\,\d \q=
\frac{
{\mkE}
\,{{\mathrm e}^{-{\eta}\,t}}
}2
,
\label{FAST-asymptote-stage1}
\\
\begin{multlined}	
\hat\Pi_0^{(1)}
=
{\mkE}
\left(
\frac{
t^{-3/2}}{8\sqrt{2\pi\eta}\,\omega_0}
+
\frac{t^{-5/2}\big(
%(-12n^2+
3
%)
\eta^2+36\omega_0^2\big)\sqrt2}{512\sqrt\pi\,\eta^{3/2}\omega_0^3}
\right)
\\+O(t^{-7/2})
+O\bigg(\frac {\mathrm e^{-\eta t}}{t}\bigg),
\end{multlined}
\label{FAST-Pi2n}
\\
    \Pi_0^{(2)} = -\frac{4
    %(-1)^n
    {\mkE}
    \omega_0
    \mathrm e^{-\eta t}
    }{\sqrt{2\pi t}
    8\omega^{3/4}
    }\,
    \cos\Big(4\omega t-\frac\pi4\Big)
+O\bigg(\frac {\mathrm e^{-\eta t}}{t}\bigg).
\label{FAST-as-pifinal}
\end{gather}
Here symbol $\PV$ means the Cauchy principal value for 
the corresponding improper integral.
%%%Integral 
%%%$\hat\Pi_0^{(0)}$ can be calculated  in the closed form 
%%%(see Appendix~\ref{FAST-App-Pi0})
%%%\begin{equation}
%%%\hat\Pi_0^{(0)}=
%%%\end{equation}
%%%%%%At the same time the $R_0(\epsilon,t)$ converges to zero
%%%%%%\begin{equation}
%%%%%%R_0(\epsilon,t)\rightrightarrows 0,\qquad{\epsilon\to+0}
%%%%%%\end{equation}
%%%%%%{uniformly in} $t$ (see
%%%%%%Appendix~\ref{FAST-App-uni}).
%%%
%%%
%%%
%%%In the particular case $\eta=0$ 
%%%the integral $\hat\Pi_{n}^{(1)}$ is zero since the integration is carried out
%%%over the interval of zero length.

\subsection{The total energy $\hat{\mathscr E}_0$}
Calculating the asymptotics by formulas
Calculating the asymptotics by Eqs.~\eqref{FAST-lagrangian-def},
\eqref{FAST-E-def},
\eqref{FAST-DTFT-inversion-rep},
\eqref{FAST-DTFT-inversion-Pi-rep}
results in:
\begin{gather}
\hat{\mathscr E}_n=
2\hat{\Pi}_n^{(0)}
+
\hat{\mathscr E}_n^{(1)}
+
\hat{\mathscr E}_n^{(2)},\\
\hat{\mathscr E}_n^{(1)}=\hat\Lambda_n^{(1)}
+2\hat\Pi_n^{(1)}
,\\
\hat{\mathscr E}_n^{(2)}=\hat\Lambda_n^{(2)}
+2\hat\Pi_n^{(2)}.
\end{gather}%\pagebreak[4]
For the non-oscillating term 
$\hat{\mathscr E}_0^{(1)}$ 
due to Eqs.~\eqref{FAST-l2n},
\eqref{FAST-Pi2n}
one has
\begin{multline}
\hat{\mathscr E}_0^{(1)}
=
{\mkE}
\left(
\frac{
t^{-3/2}}{8\sqrt{2\pi\eta}\,\omega_0}
+
\frac{t^{-5/2}\Big(
%\big(n^2-\frac14\big)
\eta^2+20\omega_0^2\Big)3\sqrt2}{512\sqrt\pi\,\eta^{3/2}\omega_0^3}
\right)
\\+O(t^{-7/2})
+O\bigg(\frac {\mathrm e^{-\eta t}}{t}\bigg).
\label{FAST-E1n}
\end{multline}
For the oscillating term 
$\hat{\mathscr E}_0^{(2)}$ 
due to 
\eqref{FAST-as-final},
\eqref{FAST-as-pifinal}
one gets 
\begin{multline}
    \hat{\mathscr E}_0^{(2)} =
%\frac{{\mkE}(-1)^n
%    \mathrm e^{-\eta t}
%}{\sqrt{2\pi t}}
%    \Bigg(
%    \Bigg(
%    \frac{2\sqrt\omega}{2\omega_0}+
%    \frac{\omega_0}{
%    8\omega^{3/4}
%    }\,
%    \Bigg)
%    \cos\Big(4\omega t-\frac\pi4\Big)
%    \\
%    -\frac{\eta}{2\omega_02\sqrt\omega}
%    \sin\Big(4\omega t-\frac\pi4\Big)
%    \Bigg)
%+O\bigg(\frac {\mathrm e^{-\eta t}}{t}\bigg)
%\\=
\frac{{\mkE}
%(-1)^n
    \mathrm e^{-\eta t}
}{2\omega_0\sqrt{2\pi t}}
    \Bigg(
    \frac{-\eta^2}
    {8\omega^{3/4}
    }\,
    \cos\Big(4\omega t-\frac\pi4\Big)
    \\
    -\frac{\eta}{2\sqrt\omega}
    \sin\Big(4\omega t-\frac\pi4\Big)
    \Bigg)
+O\bigg(\frac {\mathrm e^{-\eta t}}{t}\bigg).
\label{FAST-E2-final}
\end{multline}
The latter term becomes zero as $\eta\to+0$.

\subsection{The kinetic energy $\hat{\mathscr K}_0$}
Calculating the asymptotics by Eqs.~\eqref{FAST-lagrangian-def},
\eqref{FAST-DTFT-inversion-rep},
\eqref{FAST-DTFT-inversion-Pi-rep}
results in:
\begin{gather}
\hat{\mathscr K}_n=
\hat{\Pi}_n^{(0)}
+
\hat{\mathscr K}_n^{(1)}
+
\hat{\mathscr K}_n^{(2)},\\
\hat{\mathscr K}_n^{(1)}=\hat\Lambda_n^{(1)}
+\hat\Pi_n^{(1)}
,\\
\hat{\mathscr K}_n^{(2)}=\hat\Lambda_n^{(2)}
+\hat\Pi_n^{(2)}.
\end{gather}

For the non-oscillating term 
$\hat{\mathscr K}_n^{(1)}$ 
due to Eqs.~\eqref{FAST-l2n},
\eqref{FAST-Pi2n}
one has
\begin{equation}
\hat{\mathscr K}_n^{(1)}
=
{\mkE}\,
\frac{3\sqrt2}{64\sqrt\pi\,\omega_0\eta^{3/2}}\,
t^{-5/2}
+O(t^{-7/2})
+O\bigg(\frac {\mathrm e^{-\eta t}}{t}\bigg)
\label{FAST-as-K1}
\end{equation}
for any integer $n$.
Note that the principal term of expansion for
$\hat{\mathscr K}_n^{(1)}$ is of order $t^{-5/2}$ and does not depend on $n$, whereas expansions for
$\hat{\Pi}_n^{(1)}$, $\hat{\Lambda}_n^{(1)}$, $\hat{\mathscr E}_n^{(1)}$ have principal terms of order 
$t^{-3/2}$.
%that depend\footnote{See Eqs.~\eqref{FAST-la1-app}, \eqref{FAST-Pi1-app}}
%on $n$.

For the oscillating term 
$\hat{\mathscr K}_0^{(2)}$ 
due to 
\eqref{FAST-as-final},
\eqref{FAST-as-pifinal}
one gets 
\begin{multline}
    \hat{\mathscr K}_0^{(2)} =
%\frac{{\mkE}(-1)^n
%    \mathrm e^{-\eta t}
%}{\sqrt{2\pi t}}
%    \Bigg(
%    \Bigg(
%    \frac{2\sqrt\omega}{2\omega_0}+
%    \frac{\omega_0}{
%    8\omega^{3/4}
%    }\,
%    \Bigg)
%    \cos\Big(4\omega t-\frac\pi4\Big)
%    \\
%    -\frac{\eta}{2\omega_02\sqrt\omega}
%    \sin\Big(4\omega t-\frac\pi4\Big)
%    \Bigg)
%+O\bigg(\frac {\mathrm e^{-\eta t}}{t}\bigg)
%\\=
\frac{{\mkE}
%(-1)^n
    \mathrm e^{-\eta t}
}{2\omega_0\sqrt{2\pi t}}
    \Bigg(
    \frac{8\omega_0^2-\eta^2}
    {8\omega^{3/4}
    }\,
    \cos\Big(4\omega t-\frac\pi4\Big)
    \\
    -\frac{\eta}{2\sqrt\omega}
    \sin\Big(4\omega t-\frac\pi4\Big)
    \Bigg)
+O\bigg(\frac {\mathrm e^{-\eta t}}{t}\bigg).
\label{FAST-K2-final}
\end{multline}

\subsection{The conservative case ($\eta=0$)}
In the particular case $\eta=0$ 
the terms of asymptotic expansions for energetic quantities with superscript 
``(1)'' equal zero since the integration 
is carried out
over the interval of zero length. The corresponding asymptotic formulas for
these terms 
are not valid.
At the same time, according to 
Eqs.~\eqref{FAST-Pi0n-int}, \eqref{FAST-Pi0F-exprapp}
\begin{equation}
\hat\Pi_0^{(0)}=
\frac{\mkE}2.
\end{equation}
This yields the same results as ones obtained in 
Section~\ref{FAST-Sec-conservative} by a different approach.

\subsection{The critically damped and the overdamped cases %($\eta>4\omega_0$)
}
\label{FAST-Sec-overdamped}
In the critically damped ($\eta=4\omega_0$) and the overdamped cases ($\eta>4\omega_0$) 
the terms of asymptotic expansions for energetic quantities with superscript 
``(2)'' equal zero since the integration 
is carried out over the interval of zero length. All other formulas remain
valid.

%\subsection{Domain of applicability for asymptotic formulas}

\section{Numerics}
\label{FAST-sec-numerics}
In this section, we present the results of the numerical solution of the system of
ordinary differential equations \eqref{FAST-1}
with random initial conditions
\eqref{FAST-ic-stochastic-kao1},
\eqref{FAST-ic-stochastic-kao2}. 
It is useful to rewrite Eqs.~\eqref{FAST-1}
in the dimensionless form
\begin{equation}
\dt {\tilde{v}}_i =  %(
\L_{i} {\tilde{u}}_i - \eta {\tilde{v}}_i
%)
%d \tilde t
,
%+
%\tilde b_i \rho_i \sqrt{d \tilde t},\\ 
\qquad
\dt {\tilde{u}}_i= {\tilde{v}}_i ,%d\tilde t, 
\label{FAST-dimless}
\end{equation}
{where}
\begin{equation}
\tilde u\=\frac u a,\quad \tilde v\=\frac v {\omega_0 a},\quad \tilde t\={\omega_0} t,
%\quad \tilde b\=\frac b{c\sqrt{\omega_0}},
\quad \tilde\eta\=\frac\eta{\omega_0}.
\label{FAST-all-dless}
\end{equation}
Here $a$ is a constant with dimension of length, e.g., the lattice constant
(the distance between neighboring particles) \cite{gavrilov2018heat,gavrilov2019steady}.
We consider the chain of $2N+1$ particles and the
periodic boundary conditions
\begin{equation}
%\begin{aligned}
    u_{-N} = u_{N},  \qquad v_{-N} = v_N.
%    \\
%    v_{-N} &= v_{N-1},  &\qquad &v_{-N+1} = v_N.
%\end{aligned}
\label{FAST-bc-periodic}
\end{equation}
The initial conditions that correspond to 
\eqref{FAST-ic-stochastic-kao1} are 
\begin{gather}
\tilde u_i(0) =0,\qquad 
\tilde v_i(0)=\rho_i,
\label{FAST-ic-stochastic-kao1-dless}
\end{gather}
where 
$\rho_i$ are generated normal random numbers that satisfy 
\eqref{FAST-ic-stochastic-kao2}, where,
without loss of generality, we can take $\kappa_{i}(0)=1$.
We use 
{\sc SciPy} software \cite{scipy}: the numerical solutions of system of ODE
\eqref{FAST-dimless} are found using the standard {\sc Python} routine {\tt
scipy.integrate.odeint}.
We perform a series of $r=1\dots R$ realizations of these calculations (with
various independent 
$\rho_{(r)i}$) and get the corresponding displacements and particle
velocities as functions of discrete time $t^j$:
$\tilde u_{(r)i}(t^j)$ and  
$\tilde v_{(r)i}(t^j)$, respectively, where 
$t=\tilde t^j\=j\Delta \tilde t$.

%To obtain a numerical solution in the case of the point source of the heat
%supply located at $i=0$, we assume that 
%$\tilde b_i\rho_i=\delta_{i0}\tilde b\rho_i$ and
%use the scheme 
%\begin{equation}
%\begin{aligned}	
%\Delta {\tilde{v}}_i^j &=  (\L_{i} {\tilde{u}}_i^j - \eta {\tilde{v}}_i^j)
%\Delta {\tilde{t}} +
%\tilde b\delta_{i0} \rho^j \sqrt{\Delta {\tilde{t}}},\\ 
%\Delta {\tilde{u}}_i^j &= {\tilde{v}}_i^{j+1} \Delta {\tilde{t}},
%\\
%\tilde v^{j+1}_{i} &= \tilde v^{j}_{i} + \Delta \tilde v^{j}_{i}
%,
%\\
%\tilde u^{j+1}_{i} &= \tilde u^{j}_{i} + \Delta \tilde u^{j}_{i},
%\end{aligned}
%\label{FAST-scheme}
%\end{equation}
%where  $i=\overline{-N, N}$. 
%Here the symbols  with
%superscript $j$ denote the corresponding quantities at $\tilde
%$\rho^j$ are normal random numbers that satisfy \eqref{FAST-82} generated for all
%$\tilde t^j$.  Without loss of generality we can take $\tilde b=1$.
%
%We perform a series of $r=1\dots R$ realizations of these calculations (with
%various independent 
%$\rho^j_{(r)}$) and get the corresponding particle velocities $v^j_{i(r)}$. 

%In
%accordance with
%\eqref{FAST-temp-def},    
According to Eqs.~\eqref{FAST-4},
\eqref{FAST-genkin},
\eqref{FAST-Pi-calc-formula}, and
\eqref{FAST-mkE-def}
the ratios 
$\tilde {\mathscr K}_0{\,}_i$
and
$\tilde {\Pi}_0{\,}_i$
of the dimensionless kinetic 
and potential energies 
to the initial value of the total energy
can be
calculated as the following averages:
\begin{gather}
\tilde {\mathscr K}_0{\,}_i =\frac{1}{R}\sum_{r=1}^{R} (v_{(r)i})^2,
\label{FAST-kin-temp-num}
\\
\tilde {\Pi}_0{\,}_i =-\frac{1}{R}\sum_{r=1}^{R} 
\left(
u_{(r)i}u_{(r)i-1}
+
u_{(r)i}u_{(r)i+1}
-2(u_{(r)i})^2
\right),
\label{FAST-Pi-calc-formula-dless}
\end{gather}
respectively. 
The ratios 
$\tilde {\Lambda}_{0\,i}$
and
$\tilde {\mathscr E}_0{\,}_i$
of the dimensionless Lagrangian 
and the total energy 
to the initial value of the dimensionless total energy are
\begin{gather}
\tilde\la_0\equiv\tilde{\mathscr K}_0-\tilde\Pi_0,
\label{FAST-lagrangian-def-dless}
\\
\tilde{\mathscr E}_0\equiv
\tilde{\mathscr K}_0+\tilde\Pi_0.
\label{FAST-E-def-dless}
\end{gather}
All calculations were performed for the following values of the problem
parameters: $N=20$, $R=10000$. We verify that the numerical results 
for 
$\tilde {\mathscr K}_0{\,}_i$ and 
$\tilde {\Pi}_0{\,}_i$ 
actually do not depend on $i$, and use quantities 
$\tilde {\mathscr K}_0\equiv \tilde {\mathscr K}_0{\,}_0$ and 
$\tilde {\Pi}_0\equiv \tilde {\Pi}_0{\,}_0$ to compare the numerical and
analytical results.

\begin{figure}[htbp]	
\centering\includegraphics[width=\columnwidth]{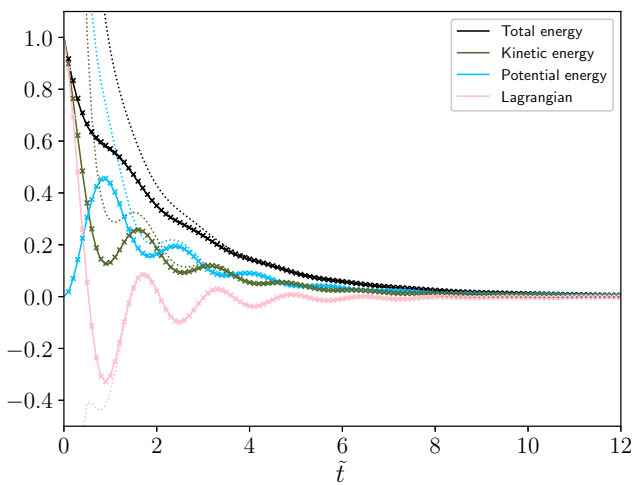}
\caption{The ratios of the total energy 
$\tilde{\mathscr E}_0$, %shown by the magenta color), 
the kinetic energy $\tilde{\mathscr K}_0$, % shown by the green color),
the potential energy  $\tilde{\Pi}_0$, % shown by the blue color),
and the Lagrangian  $\tilde{\Lambda}_0$ % shown by the red color) 
to the
initial value of the total energy versus the time $\tilde t$ in the underdamped case
($\tilde\eta=0.5$).
Comparing the analytical solutions in the integral form  
given by Eqs.~\eqref{FAST-E-calc},
\eqref{FAST-K-calc},
\eqref{FAST-DTFT-inversion-Pi-l0},
\eqref{FAST-DTFT-inversion-l0}, respectively
(the solid lines);
the corresponding numerical solutions (the crosses); and the corresponding 
asymptotic solutions (the dotted lines)}
\label{fast_time.eps}
\end{figure}

\begin{figure}[htbp]	 
\centering\includegraphics[width=\columnwidth]{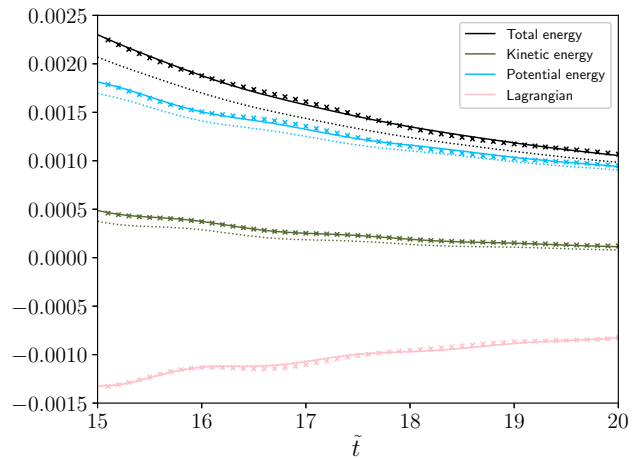}
\caption{The same quantities as in Figure~\ref{fast_time.eps} versus the time $\tilde t$
for large values of time (the underdamped case $\tilde\eta=0.5$)}
\label{fast_long.eps}
\end{figure}

\begin{figure}[htbp]	 
\centering\includegraphics[width=\columnwidth]{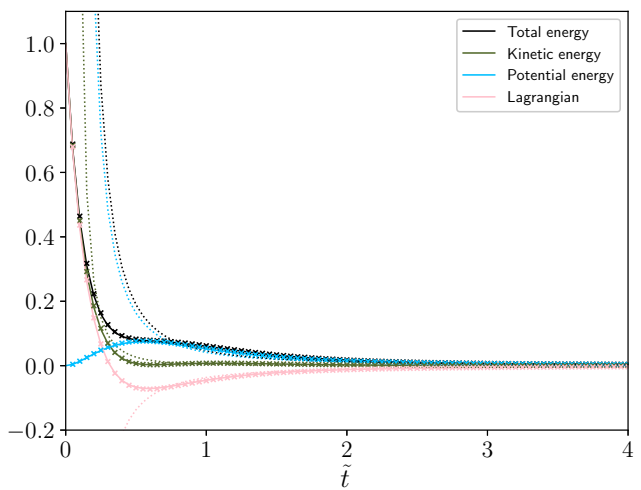}
\caption{The same quantities as in Figure~\ref{fast_time.eps} versus the time $\tilde t$
in the critically damped case ($\tilde\eta=4$)}
\label{fast_critical.eps}
\end{figure}

\begin{figure}[htbp]	 
\centering\includegraphics[width=\columnwidth]{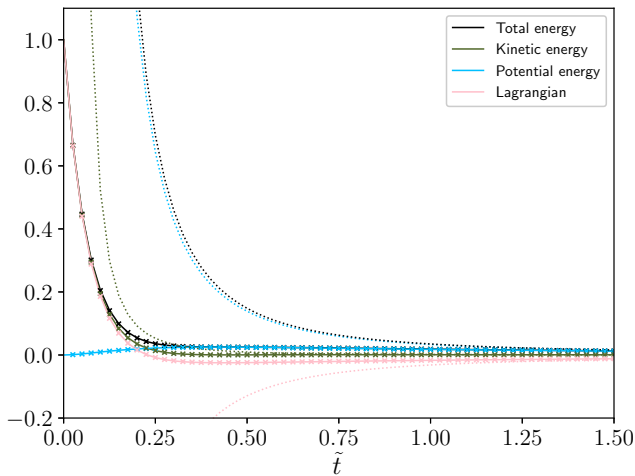}
\caption{The same quantities as in Figure~\ref{fast_time.eps} versus the time $\tilde t$
in the overdamped case ($\tilde\eta=8.0$)}
\label{fast_overdamped.eps}
\end{figure}

Numerical results  for
$\tilde{\mathscr E}_0,$
$\tilde{\mathscr K}_0,$
$\tilde{\Pi}_0,$
$\tilde{\Lambda}_0$
can be compared with
the analytical solutions in the integral form  
given by Eqs.~\eqref{FAST-E-calc},
\eqref{FAST-K-calc},
\eqref{FAST-DTFT-inversion-Pi-l0},
\eqref{FAST-DTFT-inversion-l0}, respectively,
and corresponding asymptotics (see formulas in Section~\ref{FAST-section-AS}).
The analytical solutions in the integral form are calculated using the
standard {\sc Python}
routine {\tt scipy.integrate.quad}.
A comparison in the case $\tilde\eta=0.5$ 
(the underdamped case)
is presented 
in Figures~\ref{fast_time.eps}--\ref{fast_long.eps}. 
Figure~\ref{fast_long.eps} corresponds to the case of large values of time,
when exponentially-decaying terms of the asymptotics almost vanish.
One can see that the numerical and
analytical solutions in the integral form are in a very good agreement.
Also, the same graphs confirm the accuracy of formulas for the asymptotic
solution for large values of time presented in Section~\ref{FAST-section-AS}. 
The analogous comparisons for the critically damped and overdamped cases are given in
Figures~\ref{fast_critical.eps},\ref{fast_overdamped.eps}, respectively.

%\TODO{We need to mark axes on Figures!}
%\begin{figure}[p]	
%\centering\includegraphics[width=\textwidth]{omega1_0_eta0_01_n1000_T70_0_R10000.eps}
%\caption{Comparison between the unsteady analytical solution 
%\eqref{result-nonst}
%for the crystal in a viscous environment (the green
%dashed line), the corresponding steady-state analytical solution 
%\eqref{basic-result}
%(the blue solid line) and the 
%numerical one (the red crosses)
%The parameters are the same as in
%Fig.~\ref{F1-non}, but $R=10000$.}
%\label{F2-non}
%\end{figure}

%Formulas \eqref{FAST-as-final}, \eqref{FAST-l2n}, \eqref{FAST-ln} are in very good
%agreement with numerical calculations of integral 
%\eqref{DTFT-inversion} for large $t$. \DANGER{Despite that
%$\Lambda_n^{(1)}=o(\Lambda_n^{(2)})$ in practical calculations one can take
%$\Lambda_n\simeq\Lambda_n^{(1)}$ for small $\eta$, and 
%$\Lambda_n\simeq\Lambda_n^{(2)}$ for enough big $\eta$}.
%
%
%\selectlanguage{russian}
%%Тогда для решения исходного уравнения \eqref{f6} имеем
%%\be{f15}
%%    \la_n = \si^2e^{-\eta t}J_{2n}(4\w_0 t).
%%\ee
%Таким образом, лагранжиан стремиться к нулю, а, значит, кинетическая энергия становится равной потенциальной и половине полной энергии. При этом, на колебательный степенной характер затухания накладывается слабое экспоненциальное затухание, мало проявляющееся при малых временах, однако, играющее доминирующую роль при больших временах.
%\selectlanguage{english}

\section{Discussion}
\label{FAST-Sec-discussion}
In this section we discuss the large-time behavior of energetic quantities in
the underdamped case.

From the formal point of view, the exponentially decaying terms of the
asymptotic expansions (the ones with superscripts ``$(0)$'' and ``$(2)$'') are
much less than the power-decaying terms (the ones with superscript ``$(1)$'').
It seems, therefore, that the exponentially decaying terms can be dropped out. 
However, the calculations show that actually the terms with different decay
approximate the solution at different timescales.

In Figures~\ref{fast_time.eps}, \ref{fast_long.eps} one can see that
in the underdamped case %\footnote{%
%In the paper, it is generally assumed that the underdamped case is under
%consideration, where restriction
%\eqref{FAST-eta-restriction} is fulfilled}
the unsteady process of thermal equilibration has two stages. At the first
stage (``the large times''), when the quantity 
$
{{\mathrm e}^{-{\eta}\,t}}
$
is not yet very small,
the {\it qualitative} description of the process is as follows: the kinetic
and potential energies oscillate approaching the curvilinear asymptote. The 
asymptote corresponds to the term 
$\hat\Pi_0^{(0)}$
described by formula
\eqref{FAST-asymptote-stage1}, which is equal to
$
\frac{
{\mkE}
\,{{\mathrm e}^{-{\eta}\,t}}
}2
$. 
The total energy also oscillates (with a smaller amplitude than
the kinetic and potential energy) approaching the
asymptotic level 
$
{\mkE}
\,{{\mathrm e}^{-{\eta}\,t}}
$. 
The Lagrangian oscillates around zero and approaches zero. The oscillatory
motions are described by the terms
$\tilde{\mathscr E}_0^{(2)},$
$\tilde{\mathscr K}_0^{(2)},$
$\tilde{\Pi}_0^{(2)},$
$\tilde{\Lambda}_0^{(2)}$
expressed by formulas 
\eqref{FAST-E2-final},
\eqref{FAST-K2-final},
\eqref{FAST-as-pifinal},
\eqref{FAST-as-final}, respectively.
The amplitudes of oscillations for all energetic quantities are of order 
$O\left(t^{-1/2}{{\mathrm e}^{-{\eta}\,t}}\right)$.  In the limiting case of
zero dissipation the first stage transforms into the solution describing the
thermal equilibration in the corresponding conservative system. 

%Note that 

Unexpectedly, there is the second stage (see~Figure~\ref{fast_long.eps}), that
can be observed
%(in the case of the small enough specific viscosity)
only when the quantity 
$
{{\mathrm e}^{-{\eta}\,t}}
$
becomes very small \footnote{The stage where the asymptotics have a power
decay exists for any positive value of
specific viscosity $\eta$} (``the very large times'').
The expressions for all energetic quantities at the
second stage are subjected to a power decay, i.e.\ 
from the formal point of view the corresponding terms 
$\tilde{\mathscr E}_0^{(1)},$
$\tilde{\mathscr K}_0^{(1)},$
$\tilde{\Pi}_0^{(1)},$
$\tilde{\Lambda}_0^{(1)}$
are principal terms of the corresponding asymptotic expansions. The formulas for these
terms are 
\eqref{FAST-E1n},
\eqref{FAST-as-K1},
\eqref{FAST-Pi2n},
\eqref{FAST-l2n}, respectively. Another one unexpected result is as follows:
the principal term of the asymptotic expansion for 
$\tilde{\mathscr K}_0$ is proportional to $t^{-5/2}$ in the case
of the kinetic energy and to $t^{-3/2}$ for all other energetic quantities. 
In the limiting case of
zero dissipation the second stage disappears.

\begin{figure}[htbp]	
\centering\includegraphics[width=\columnwidth]{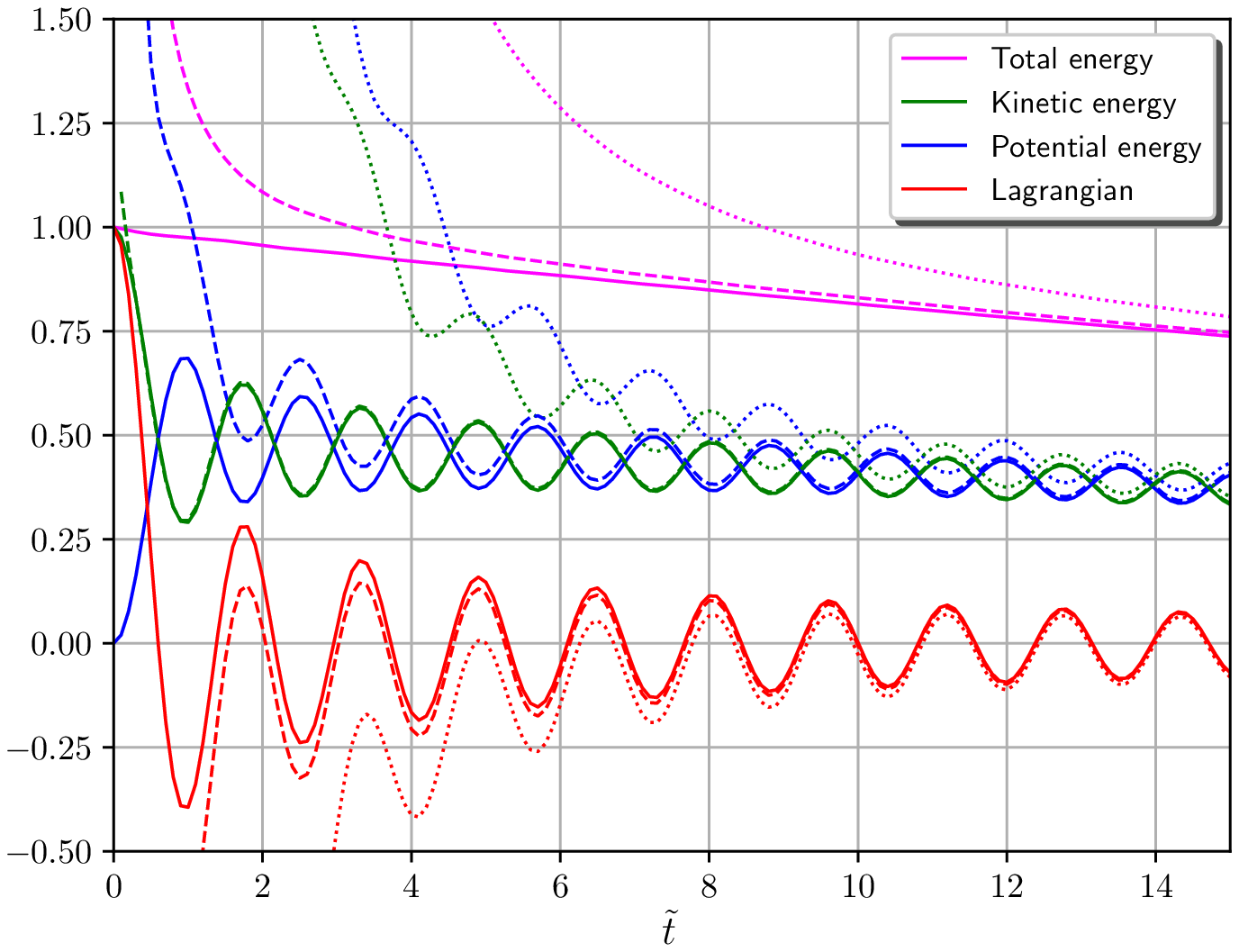}
\caption{The ratios of the total energy 
$\tilde{\mathscr E}_0$, %shown by the magenta color), 
the kinetic energy $\tilde{\mathscr K}_0$, % shown by the green color),
the potential energy  $\tilde{\Pi}_0$, % shown by the blue color),
and the Lagrangian  $\tilde{\Lambda}_0$ % shown by the red color) 
to the
initial value of the total energy versus the time $\tilde t$ in the case of very small
dissipation 
($\tilde\eta=0.02$).
Comparing the analytical solutions in the integral form  
given by Eqs.~\eqref{FAST-E-calc},
\eqref{FAST-K-calc},
\eqref{FAST-DTFT-inversion-Pi-l0},
\eqref{FAST-DTFT-inversion-l0}, respectively
(the solid lines);
the corresponding 
asymptotic solutions (the dotted lines), and approximate asymptotic solution,
wherein the terms with superscripts ``(1)'' are dropped out (the dashed lines)}
\label{fast_smalleta.eps}
\end{figure}

The calculations show that in the case $\tilde \eta\ll1$ the asymptotic
formulas for the power-decaying terms %(the ones with superscripts ``$(1)$'')
can give wrong results at a timescale that corresponds to the large, but not very large times.
Thus, to approximate the solution at such a timescale,
it can be preferable to drop out the power-decaying terms 
and use only the exponentially decaying terms of
asymptotics (the ones with superscripts ``$(0)$'' and ``$(2)$'').
This fact
is illustrated in Figure~\ref{fast_smalleta.eps}.

\section{Conclusion}
\label{FAST-Sec-conclusion}
It the paper, we have obtained 
the analytical solutions 
in the integral form 
describing the thermal equilibration in a one-dimensional damped harmonic
crystal
(see Section~\ref{FAST-Sec-analyt-intform}).
These solutions describe the time behavior of the energetic quantities
(the total energy, the kinetic energy, the potential
energy, and the Lagrangian).
The analytical solutions are in an excellent agreement with
independent numerical calculations 
(see Section~\ref{FAST-sec-numerics}).

The most important result of the paper is large-time asymptotic formulas
for energetic quantities  presented in 
Section~\ref{FAST-section-AS} (see the derivation in the 
Appendices~\ref{FAST-App-L2}--\ref{FAST-App-P0}).

The main conclusions of the paper can be formulated as follows:
\begin{itemize}	
\item 
In presence of small viscous external damping (i.e., in the underdamped case)
the process of thermal equilibration 
is more complicated than in the corresponding conservative system and has two
stages that correspond to large and very large times
(see Figures~\ref{fast_time.eps}, \ref{fast_long.eps}).
\item
In
the critically damped 
(see Figure~\ref{fast_critical.eps})
and the overdamped 
(see Figure~\ref{fast_overdamped.eps})
cases the process of thermal equilibration has only one stage, which is
similar to the second stage in the underdamped case.
\item
At very large times (i.e., during the second stage) the total energy of an
underdamped harmonic crystal is mostly the potential energy. The same
conclusion is true for large time behavior of a critically damped and an
overdamped harmonic crystal.
\end{itemize}

\section*{Acknowledgments}
The authors are grateful to V.A.~Kuzkin and E.V.~Shishkina for useful and
stimulating discussions. This work is supported by Russian Science Foundation (Grant No. 
19-41-04106).

%than the corresponding terms with superscript
% subjected to power decay, and therefore can be dropped out. The
%calculations show that in reality these terms approximate the solution at
%different timescales.

%\begin{subappendices}	

\appendix

\section{Calculation of the asymptotics for $\hat\la_n^{(2)}$}
\label{FAST-App-L2}

Following to the general procedure of the method of stationary phase \cite{Fedoruk-Saddle},
according to the localization principle, we claim that for $t\to\infty$ the integral 
$\hat\Lambda_{n}^{(2)}$ defined by
Eqs.~\eqref{FAST-DTFT-inversion-rep},
\eqref{FAST-lambdaF-e}
equals the sum 
\begin{equation}
\hat\Lambda_n^{(2)}=\sum_i\tilde\Lambda_n^{(2)}(\q_i)+O(t^{-\infty})
\end{equation}
of contributions 
$\tilde\Lambda_n^{(2)}(\q_i)$
from the critical points $\q=\q_i$
%\footnote{}
in the
interval of integration $[\bar \q;\pi]$. 
In the case under consideration, there are
two critical points: the stationary point $\q_1=\pi$ 
of the phase function
\begin{equation}
\phi(\q)=\sqrt{|\mathscr A^2(\q)-\eta^2|},
\label{FAST-app-psipm}
\end{equation}
and the boundary of
integration interval $\q_2=\bar \q$.
%
%{One can show} that the principal term of asymptotic expansion for
%large times of the integral $\Lambda_n^{(2)}$ is
The contribution from the stationary point $\q_1=\pi$ can be calculated as 
\begin{multline}
\tilde\Lambda_n^{(2)}(\q_1)
=\frac{{\mkE}{\mathrm
e}^{-{\eta}\,t}}{\pi}
\int_{\bar \q}^{\pi}
\chi(\q-\pi)
\left(
\cos \big( \sqrt {\mathscr A^2-{{\eta}}^{2}}
\,t
 \big)
\right.
\\
\left.
-\frac {{\eta}\,\sin  \big( \sqrt {\mathscr A^2-\eta^2}\,t \big) }
        {\sqrt {\mathscr A^2-\eta^2}}
\right)
\cos n\q\,\d \q.
%\\+O\bigg(\frac {\mathrm e^{-\eta t}}{t}\bigg).
\label{FAST-contribute-fast-eta}
\end{multline}
Here and in what follows,
$\chi(\q)\in C^\infty$ is a non-negative even function such that
\begin{equation}
\begin{aligned}	
&\chi(\q)\equiv 1\quad &\text{for} \quad
&|\q|<\frac{\bar\epsilon}3,
\\
&\chi(\q)\equiv 0\quad &\text{for} \quad
&|\q|>\frac{2\bar\epsilon}3;
\end{aligned}
\label{FAST-chi-def}
\end{equation}
$\bar\epsilon>0$ is a small enough number to get the integrand with a unique isolated 
critical point.
The expression for the principal term of the asymptotics for the contribution
from an isolated boundary stationary point
$\q=\q_\ast$ is \cite{Fedoruk-Saddle}
\begin{multline}
\int_{-\infty}^{\q_\ast} \chi(\q-\q_\ast)f(\q)\,\exp(\I\phi(\q)t)\, \d \q\\=
\left(\frac12\sqrt{\frac{2\pi}{|\phi''(\q_\ast)|t}}\,
f(\q_\ast)+O(t^{-1})\right)
\\\times
\exp\left(\I\phi(\q_\ast)t+\frac{\I\pi}4\sign\phi''(\q_\ast)\right).
%+O(t^{-3/2}).
\label{FAST-1:MSP-principal-part}
\end{multline}
%For phase function $\phi$ is 
%\begin{equation}
%\phi(q)=\sqrt{\mathscr A^2(q)-\eta^2},
%\end{equation}
%and the value of the phase function and its second derivative at the
%stationary point are
One has
\begin{gather}
\phi(\pi)=\sqrt{16\omega_0^2-\eta^2},
\label{FAST-phipi}
\\
\phi''(\pi)=-\frac{4\omega_0^2}{\sqrt{16\omega_0^2-\eta^2}}.
\label{FAST-phipidd}
\end{gather}
Taking the real part of formula
\eqref{FAST-1:MSP-principal-part},
wherein $f(\q)=\lambda(\q)$,
\begin{equation}
\lambda(\q)\equiv
\frac{{\mkE}{\mathrm e}^{-{\eta}\,t}}{\pi}
\left(1+\frac{\I\eta}{\sqrt{\mathscr A^2(\q)-\eta^2}}\right)\,\cos n\q
\end{equation}
to calculate the integral in the right-hand side of 
Eq.~\eqref{FAST-contribute-fast-eta}
results in
%\TODO{\foreignlanguage{russian}{Необходима численная проверка! Очень много
%нюансов!}}
\begin{multline}
    \tilde\la_n^{(2)}(\pi) =
    {\mkE}(-1)^n\frac{\sqrt[4]{16\omega_0^2-\eta^2}}{2\omega_0\sqrt{2\pi t}}\,
    \mathrm e^{-\eta t}
    \\\times
    \Bigg(
    \cos\Big(\sqrt{16\omega_0^2-\eta^2}\, t-\frac\pi4\Big)
    \\
    -\frac{\eta}{\sqrt{16\omega_0^2-\eta^2}}
    \sin\Big(\sqrt{16\omega_0^2-\eta^2}\, t-\frac\pi4\Big)
    \Bigg)
+O\bigg(\frac {\mathrm e^{-\eta t}}{t}\bigg).
\label{FAST-as-final-app}
\end{multline}

To finalize the calculation of asymptotics for $\hat\la_n^{(2)}$ we need to
estimate the contribution from the critical point $\q_2=\bar \q$. One has 
\begin{gather}
\phi(\q)\sim \phi_{1/2}|\q-\bar \q|^{1/2}+\phi_{3/2}|\q-\bar \q|^{3/2}+\dots, 
%\quad \q\to\bar \q+0;
\label{FAST-app-phi}
\\
\Im \lambda(\q)\sim 
\frac{{\mkE}\eta\,{\mathrm e}^{-{\eta}\,t}}{\phi_{1/2}\pi}
 |\q-\bar \q|^{-1/2}+\dots,
\label{FAST-phi-approx}
\end{gather}
as $\q\to\bar \q+0$. Here $\phi_{1/2}$ is a positive constant. 
It follows from the Erdeliy lemma
\cite{Fedoruk-Saddle} that
\begin{equation}
\int_0^\infty \q^{\beta-1}\chi(\q)\exp(\I t \q^\alpha)\,\d \q\sim
A t^{-\frac{\beta}{\alpha}},
\label{FAST:Erdelyi}
\end{equation}
where $A$ is a non-zero constant. Applying Eq.~\eqref{FAST:Erdelyi} in the cases
$\beta=1,\ \alpha=1/2$ and $\beta=1/2,\ \alpha=1/2$
yields
\begin{equation}
\tilde\la_n^{(2)}(\bar \q) = 
O\bigg(\frac {\mathrm e^{-\eta t}}{t}\bigg)
+O\bigg(\frac {\mathrm e^{-\eta t}}{t^{2}}\bigg)=
O\bigg(\frac {\mathrm e^{-\eta t}}{t}\bigg).
\end{equation}

Thus, the asymptotics of $\hat\la_n^{(2)}$ equals to the right-hand side of 
Eq.~\eqref{FAST-as-final-app}.

\section{Calculation of the asymptotics for $\hat\la_n^{(1)}$}
\label{FAST-App-L1}

One has
\begin{equation}
\hat\Lambda_n^{(1)}
%=\frac1{2\pi}\int_{-\pi}^{\pi}\F\Lambda{\q}n\,\exp(\I n\q)\,\d \q
=
\frac1{2\pi}
\int_{-\bar \q}^{\bar \q}
\F\Lambda{\q}n\,\cos{n\q}\,\d \q
%\\=
%\frac1{\pi}
%\left(
%\int_{0}^{\bar \q}
%+
%\int_{\bar \q}^{\pi}
%\right)
%\F\Lambda{\q}n\,\cos{n\q}\,\d \q
\equiv
\hat\Lambda_n^{(1)(+)}
+
\hat\Lambda_n^{(1)(-)},
\end{equation}
where
\begin{multline}
\hat\Lambda_n^{(1)(\pm)}\equiv
\frac{{\mkE}{\mathrm e}^{-{\eta}\,t}}{2\pi}
\\\times
\int_{-\bar \q}^{\bar \q}
%\left(
\frac{ 
\big( {\mathscr A}^{2}\pm\eta\sqrt {{\eta}^{2}-{\mathscr A}^{2}}-{\eta}^{2} \big) 
{\mathrm e}^{\pm\sqrt {{\eta}^{2}-{\mathscr A}^{2}}\, t}
}{2({\mathscr A}^{2}-{{\eta}}^{2})}
%\right)
\cos n\q\,\d \q.
\end{multline}

Following to the general procedure of the Laplace method
\cite{Fedoruk-Saddle}, we claim that for $t\to\infty$ the each integral 
$\hat\Lambda_{n}^{(1)(\pm)}$ 
can be asymptotically approximated by contribution 
\begin{equation}
\hat\Lambda_n^{(2)(\pm)}\sim
\Lambda_n^{(2)(\pm)}(\q_\pm)
%\sim
%\frac{{\mkE}{\mathrm
%e}^{-{\eta}\,t}}{2\pi}p_\pm(t)
%\exp\big(\pm\phi_\pm(\q_\pm)t\big)
\end{equation}
from the global maximum point $\q_\pm$ for
the functions $\pm\phi(\q)$ (defined by 
%\begin{equation}
%\pm\phi(\q)=\pm\sqrt {|{\eta}^{2}-{\mathscr A^{2}(\q)}|}
\eqref{FAST-app-psipm})
%\end{equation}
lying in the
interval of integration $[-\bar \q;\bar \q]$. 
%Here $p_\pm(t)\to0$ are
%asymptotic expansions with structure depending on the type of the maximum point.

At first, consider the integral $\Lambda_n^{(1)(+)}$. The maximum point
for $\phi(\q)$ 
is the internal point $\q=0$, and $\phi(0)=\eta$.
The expression for the corresponding contribution
in the
case of an internal isolated maximum point $\q=\q_\star$ is
%\begin{multline}
%\hat\Lambda_n^{(1)}=
%{\mkE}
%{\mathrm e}^{-{\eta}\,t}\int_{\bar \q}^{\bar \q}
%\chi(\q)
%\left(
%\frac{ 
%\big( {\mathscr A}^{2}+\eta\sqrt {{\eta}^{2}-{\mathscr A}^{2}}-{\eta}^{2} \big) 
%{\mathrm e}^{\sqrt {{\eta}^{2}-{\mathscr A}^{2}}\, t}
%}{2({\mathscr A}^{2}-{{\eta}}^{2})}
%\right)
%\cos n\q\,\d \q
%\\+O\bigg(\frac {\mathrm e^{-\eta t}}{t}\bigg).
%\label{FAST-contribute-fast-eta-1}
%\end{multline}
%
(see \cite{Fedoruk-Saddle}, formula (1.25)
\footnote{
Note that there are misprints in (\cite{Fedoruk-Saddle}, formula (1.25)) that were
corrected in 
\eqref{FAST-L-expansion}.
}): 
\begin{multline}
\int_{-\infty}^{\infty} \chi(\q-\q_\star)f(\q)\exp\big(\phi(\q)t\big)\, \d \q
\\
\sim
%{\mkE}
\sum_{k=0}^\infty c_kt^{-k-1/2}
\exp\big(\phi(\q_\star)t\big)
%+o\bigg(\frac {\mathrm e^{-\eta t}}{t}\bigg)
,
\label{FAST-as-laplace-series}
\end{multline}
where 
\begin{multline}
c_k=\frac{\Gamma\big(k+\frac12\big)}{(2k)!}\\
\times\Big(\frac \d{\d \q}\Big)^{2k}
\Bigg(f(\q)\bigg(
\frac{\phi(\q_\star)-\phi(\q)}{(\q-\q_\star)^2}
\bigg)^{-k-\frac12}
\Bigg)\Bigg|_{\q=\q_\star}
\label{FAST-L-expansion}
.
\end{multline}
Now we take
$f(\q)=L_+(\q)$, where
\begin{gather}
L_\pm(\q)\equiv
\frac{{\mkE}{\mathrm e}^{-{\eta}\,t}}{2\pi}\,
\frac
{ 
 {\mathscr A}^{2}\pm\eta\sqrt {{\eta}^{2}-{\mathscr A}^{2}}-{\eta}^{2} 
}
{{\mathscr A}^{2}-{{\eta}}^{2}}
\cos {n\q}
%\,
%{{\mathrm e}^{-{\eta}\,t}}
,
\label{FAST-L-expansion2}
%\\
%\psi(\q)=\sqrt {{\eta}^{2}-{\mathscr A}^{2}}.
%\label{FAST-L-expansion3}
\end{gather}
to calculate the coefficients $c_k$. To do this we use {\sc Maple} symbolic calculation
software. In this way, we find 
\begin{gather}	
c_0=0,
\end{gather}
i.e. we deal with a degenerate case, and
\begin{gather}	
c_1=-\frac\mkE{8\sqrt{2\pi\eta}\,\omega_0},
\\
c_2=
\frac{\mkE\big((12n^2-3)\eta^2-12\omega_0^2\big)\sqrt2}
{512\sqrt\pi\,\eta^{3/2}\omega_0^3}.
\end{gather}
%Finally, we have 
Thus, 
\begin{multline}
\tilde\Lambda_n^{(1)(+)}=
{\mkE}\left(
-
\frac{
t^{-3/2}}{8\sqrt{2\pi\eta}\,\omega_0}
\right.
\\+
\left.
\frac{t^{-5/2}\big((12n^2-3)\eta^2-12\omega_0^2\big)\sqrt2}
{512\sqrt\pi\,\eta^{3/2}\omega_0^3}
\right)
+O(t^{-7/2}).
%+O\bigg(\frac {e^{-\eta t}}{t}\bigg).
%\label{FAST-l2n}
%\\
%\hat\Lambda_n=\hat\Lambda_n^{(1)}+\hat\Lambda_n^{(2)}.
%\label{ln}
\label{FAST-la1-app}
\end{multline}
%In the last formula we save only the principal term of order $O(t^{-3/2})$.

To finalize the calculation of asymptotics for $\hat\la_n^{(1)}$ we need to
estimate the integral $\tilde\la_n^{(1)(-)}$. The maximum points for
$-\phi(\q)$ are boundary points $\pm\bar \q$, and $-\phi(\bar \q)=0$.
Due to the symmetry, these two points
bring equal contributions, therefore we can estimate only one of them at
$\q=\bar \q$. One has
\begin{gather}
%\psi_-(\q)\sim C_1(\bar \q-\q)^{1/2},
%\label{FAST-psi-approx}
%\\
L_-(\q)\sim L_{1/2}
{\mathrm e}^{-{\eta}\,t}
(\bar \q-\q)^{-1/2},
\label{FAST-f-approx}
\end{gather}
as $\q\to\bar \q-0$. Here $L_{1/2}$ is a non-zero constant. 
It follows from the Watson lemma
\cite{Fedoruk-Saddle} that
%\TODO{$\alpha,\ \beta$}
\begin{equation}
\int_0^\infty \q^{\beta-1}\chi(\q)\exp(- t \q^\alpha)\,\d \q\sim
B t^{-\frac{\beta}{\alpha}},
\label{FAST:Watson}
\end{equation}
where $B$ is a non-zero constant. 
Taking into account 
\eqref{FAST-app-phi}
and 
using 
\eqref{FAST:Watson}
in the case 
$\beta=1/2,\ \alpha=1/2$
one gets
\begin{equation}
\tilde\la_n^{(1)(-)} = 
O\bigg(\frac {\mathrm e^{-\eta t}}{t}\bigg).
\end{equation}

Thus, the asymptotics of $\hat\la_n^{(1)}$ equals to the right-hand side of 
Eq.~\eqref{FAST-la1-app}.

\section{Calculation of the asymptotics for $\hat\Pi_n$}
\label{FAST-App-P}
%Since $\Pi_F$ is non-singular at $\q=\pm\bar \q$
%(whereas all of the quantities 
%$\Pi_F^{(0)}$,
%$\Pi_F^{(1)}$,
%$\Pi_F^{(2)}$
%have a non-integrable singularity there),
The integral in the right-hand side of 
Eq.~\eqref{FAST-DTFT-inversion-Pi} can be represented as follows:
\begin{gather}	
\hat\Pi_n=
\hat\Pi_n^{(0)}
+
\hat\Pi_n^{(1)}
+
\hat\Pi_n^{(2)}
+
R,
\label{FAST-Pi-with-R}
\\
\hat\Pi_n^{(0)}
=
\frac1\pi
\left(
\int_{0}^{\bar \q-{\epsilon}}+
\int_{\bar \q+{\epsilon}}^\pi
\right)
\Pi_F^{(0)}
\cos{n\q}\,\d \q,
\\
\hat\Pi_n^{(1)}
=
\frac1{2\pi}
\left(
\int_{-\bar \q+{\epsilon}}^{\bar \q-{\epsilon}}
\Pi_F^{(1)}
\cos{n\q}\,\d \q
\right),
\\
\hat\Pi_n^{(2)}
=
\frac1\pi
\left(
\int_{-\pi}^{-\bar \q-{\epsilon}}
+
\int_{\bar \q+{\epsilon}}^{\pi}
\right)
\Pi_F^{(2)}
\cos{n\q}\,\d \q,
%}_{\hat\Pi_n^{(2)}}
\\
\begin{multlined}	
R=
\frac1\pi
\left(
\int_{\bar \q-{\epsilon}}^{\bar \q}
\big(
\Pi_F^{(0)}
+
\Pi_F^{(1)}
\big)
\cos{n\q}\,\d \q
\right.
\\+
\left.
\int_{\bar \q}^{\bar \q+{\epsilon}}
\big(
\Pi_F^{(0)}
+
\Pi_F^{(2)}
\big)
\cos{n\q}\,\d \q
\right),
\end{multlined}
\label{FAST-reminder}
\end{gather}
where ${\epsilon}>0$ is a small enough number.
%\begin{multline}
%\hat\Pi_n
%%=\frac1{2\pi}\int_{-\pi}^{\pi}\F\Lambda{\q}n\,\exp(\I n\q)\,\d \q
%=
%%\frac1{2\pi}
%%\left(
%%\int_{-\bar \q}^{\bar \q}
%%+\int_{-\pi}^{-\bar \q}
%%+\int_{\bar \q}^\pi
%%\right)
%%\F\Lambda{\q}n\,\cos{n\q}\,\d \q
%%\\=
%%\frac1{2\pi}
%%\left(
%%\underbrace{
%\frac1\pi
%\left(
%\int_{0}^{\bar \q-{\epsilon}}+
%\int_{\bar \q+{\epsilon}}^\pi
%\right)
%\Pi_F^{(0)}
%\cos{n\q}\,\d \q
%%}_{\hat\Pi_n^{(0)}}
%\\+
%%\underbrace{
%\frac1{2\pi}
%\left(
%\int_{-\bar \q+{\epsilon}}^{\bar \q-{\epsilon}}
%\Pi_F^{(1)}
%\cos{n\q}\,\d \q
%%}_{\hat\Pi_n^{(1)}}
%+
%%\underbrace{
%%\frac1\p
%\left(
%\int_{-\pi}^{-\bar \q-{\epsilon}}
%+
%\int_{\bar \q+{\epsilon}}^{\pi}
%\right)
%\Pi_F^{(2)}
%\cos{n\q}\,\d \q
%%}_{\hat\Pi_n^{(2)}}
%\right)
%\\
%+\frac1\pi
%\left(
%\int_{\bar \q-{\epsilon}}^{\bar \q}
%\big(
%\Pi_F^{(0)}
%+
%\Pi_F^{(1)}
%\big)
%\cos{n\q}\,\d \q
%+
%\int_{\bar \q}^{\bar \q+{\epsilon}}
%\big(
%\Pi_F^{(0)}
%+
%\Pi_F^{(2)}
%\big)
%\cos{n\q}\,\d \q
%\right)
%\\
%\equiv
%\hat\Pi_n^{(0)}
%+
%\hat\Pi_n^{(1)}
%+
%\hat\Pi_n^{(2)}
%%+
%%2\lim_{{\epsilon}\to+0}R_n({\epsilon},t),
%\label{FAST-DTFT-inversion-Pi-rep}
%\end{multline}
%where symbol $\PV$ means the Cauchy principal value.  Note that 
%Eq.~\eqref{FAST-DTFT-inversion-Pi-rep} correct if and only if all quantities in the
%right-hand side exist.
%%%for the corresponding
%%%integrals with integrands singular at $\q=\bar \q$.
Again, the integral $\hat\Pi_{n}^{(1)}$ 
is a Laplace type integral,
%%%and therefore it
%%%can be estimated by the Laplace method,
whereas the
integral $\hat\Pi_{n}^{(2)}$ 
is a Fourier type integral.
%%%and therefore it can be estimated by 
%%%the method of stationary phase. 

The asymptotic expansion for
large time of the integral $\Pi_n^{(2)}$ is the sum 
of the doubled contribution from the boundary stationary point $\q=\q_\ast=\pi$
(where $\phi'(\q_\ast)=0$):
\begin{multline}
\hat\Pi_n^{(2)}(\pi)=
\\-\frac{{\mkE}{\mathrm e}^{-{\eta}\,t}}{\pi}
%\\\times
\int_{\bar \q}^{\pi}
\chi(\q-\pi)\,
%-\frac {{\eta}\,\sin  \big( \sqrt {\mathscr A^2-\eta^2}\,t \big) }
%        {\sqrt {\mathscr A^2-\eta^2}}
\frac{
\mathscr A^2\cos \big( \sqrt {\mathscr A^2-{{\eta}}^{2}}\,t
 \big)
}{\mathscr A^2-\eta^2}
\cos n\q\,\d \q
\\+O\bigg(\frac {\mathrm e^{-\eta t}}{t}\bigg),
\label{FAST-contribute-fast-eta-Pi}
\end{multline}
and the doubled contribution from the boundary point $\q=\bar \q+{\epsilon}$. 
The latter
term is discussed at the end of this Appendix
(after formula \eqref{FAST-int-cosine-codhine}), where we deal with the
estimation of the reminder $R$.
%\begin{equation}
%\hat\Pi_n^{(2)}(\bar \q+{\epsilon})=
%O\bigg(\frac {\mathrm e^{-\eta t}}{t}\bigg).
%\end{equation}
Applying formula 
\eqref{FAST-1:MSP-principal-part},
wherein $f(\q)=P(\q)$,
\begin{equation}
P(\q)\equiv
-\frac{{\mkE}{\mathrm e}^{-{\eta}\,t}\cos n\q}{\pi}
\frac{
\mathscr A^2}
%{{\mathrm e}^{-{\eta}\,t}}}
{{\mathscr A}^{2}-{{\eta}}^{2}},
\label{FAST-f(q)}
\end{equation}
to calculate the integral in the right-hand side of 
Eq.~\eqref{FAST-contribute-fast-eta-Pi}, and taking into account 
Eqs.~\eqref{FAST-phipi},
\eqref{FAST-phipidd},
results in
%Calculating the asymptotics of the last integral by formula 
%\eqref{FAST-as-final} yields
\begin{multline}
    \Pi_n^{(2)}(\pi) \\= -\frac{4(-1)^n{\mkE}
    \omega_0
    \mathrm e^{-\eta t}
    }{\sqrt{2\pi t}
    (16\omega_0^2-\eta^2)^{3/4}
    }\,
    \cos\Big(\sqrt{16\omega_0^2-\eta^2}\, t-\frac\pi4\Big)
\\+O\bigg(\frac {\mathrm e^{-\eta t}}{t}\bigg).
\label{FAST-Pi-App-as-osc}
\end{multline}

We represent integral 
$\hat\Pi_n^{(1)}$ as follows:
\begin{equation}
\hat\Pi_n^{(1)}
%=\frac1{2\pi}\int_{-\pi}^{\pi}\F\Pi{\q}n\,\exp(\I n\q)\,\d \q
=
\frac1{2\pi}
\int_{-\bar \q}^{\bar \q}
\F\Pi{\q}n\,\cos{n\q}\,\d \q
%\\=
%\frac1{\pi}
%\left(
%\int_{0}^{\bar \q}
%+
%\int_{\bar \q}^{\pi}
%\right)
%\F\Pi{\q}n\,\cos{n\q}\,\d \q
\equiv
\hat\Pi_n^{(1)(+)}
+
\hat\Pi_n^{(1)(-)},
\end{equation}
where
\begin{equation}
\hat\Pi_n^{(1)(\pm)}\equiv
-\frac{{\mkE}{\mathrm e}^{-{\eta}\,t}}{2\pi}
\int_{\bar \q}^{\bar \q}
%\left(
\frac{ 
{\mathscr A}^{2}
{\mathrm e}^{\pm\sqrt {{\eta}^{2}-{\mathscr A}^{2}}\, t}
}{4({\mathscr A}^{2}-{{\eta}}^{2})}
%\right)
\cos n\q\,\d \q.
\end{equation}
At first, consider the integral $\Pi_n^{(1)(+)}$. Again (see Appendix
~\ref{FAST-App-L1}), the maximum point
for $\phi(\q)$ defined by 
Eq.~\eqref{FAST-app-psipm}
is $\q=0$, and $\phi(0)=\eta$.
Applying formulas 
\eqref{FAST-as-laplace-series}, \eqref{FAST-L-expansion}
(wherein $f(\q)=P(\q)$, and $P(\q)$ is defined by 
\eqref{FAST-f(q)}),
%\begin{equation}
%f(\q)=
%-\frac{{\mkE}{\mathrm e}^{-{\eta}\,t}\,\cos n\q}{8\pi}
%\frac{
%\mathscr A^2\,
%}
%{{\mathscr A}^{2}-{{\eta}}^{2}}
%\end{equation}
to calculate the corresponding contribution, one can get
\begin{gather}	
c_0=0,
\end{gather}
i.e. we deal with a degenerate case, {and}
\begin{gather}	
c_1=\frac\mkE{8\sqrt{2\pi\eta}\,\omega_0},
\\
c_2=
\frac{\mkE\big((-12n^2+3)\eta^2+36\omega_0^2\big)\sqrt2}{512\sqrt\pi\,\eta^{3/2}\omega_0^3}
.
\end{gather}
Here, to calculate coefficients $c_k$, we again used {\sc Maple} symbolic calculation
software.
%
%%\section{Calculation of the asymptotics for $\hat\Pi_n^{(1)}$}
%%\eqref{FAST-App-P1}
%For the large times the integral $\Pi_n^{(1)}$ equals 
%the contribution from the maximum point $\q=\q_\ast=\pi$. Calculating the
%contribution by formulas 
%\eqref{FAST-as-laplace-series}--\eqref{FAST-L-expansion} where in 
%$\psi(\q)$ is defined by 
%\eqref{FAST-L-expansion3}, and 
%results in
In this way we find
\begin{multline}
\hat\Pi_n^{(1)(+)}
=
{\mkE}
\left(
\frac{
t^{-3/2}}{8\sqrt{2\pi\eta}\,\omega_0}
\right.
\\+
\left.
\frac{t^{-5/2}\big((-12n^2+3)\eta^2+36\omega_0^2\big)\sqrt2}{512\sqrt\pi\,\eta^{3/2}\omega_0^3}
\right)
+O(t^{-7/2}).
%+O\bigg(\frac {\mathrm e^{-\eta t}}{t}\bigg).
\label{FAST-Pi1-app}
\end{multline}

Now we need to consider the integral $\Pi_n^{(1)(-)}$. 
%Again (see Appendix
%~\ref{FAST-App-L1}), 
The maximum points
for $-\phi(\q)$ defined by 
Eq.~\eqref{FAST-app-psipm}
are boundary points $\q=\pm(\bar \q\mp{\epsilon})$:
\begin{equation}
\hat\Pi_n^{(1)}
=
\frac1{\pi}
\left(
\int_{0}^{\bar \q-{\epsilon}}
%\chi(\q-\bar \q+{\epsilon})\,
\Pi_F^{(1)}
\cos{n\q}\,\d \q
\right).
\end{equation}
The corresponding contribution is discussed in what follows
(after formula \eqref{FAST-int-cosine-codhine}), where we deal with the
estimation of the reminder term $R$.
%In this case, one has
%\begin{equation}
%\hat\Pi_n^{(1)}=O\left(
%\frac{
%\mathrm e^{\left(\psi_-(\bar \q-{\epsilon})-\eta\right)t}
%}{{\epsilon} t}
%\right)
%\end{equation}

Now consider the reminder term $R$ defined by Eq.~\eqref{FAST-reminder}.
We need to estimate the contribution to
integral 
\eqref{FAST-Pi-with-R}
from 
a neighbourhood of
the point $\q=\bar \q$.
%
%, and 
%$\psi_+(\bar \q-{\epsilon})=\psi_+(-\bar \q+{\epsilon})$.  
%At the same time $\q=\pm\bar \q$ is the
%critical point for the integral $\Pi_n^{(2)}$.
%The problem is that all quantities 
%$\Pi_F^{(1)(+)}$,
%$\Pi_F^{(1)(-)}$,
%$\Pi_F^{(2)}$
%are singular at $\pm\bar \q$. Due to the symmetry, it is enough to
%estimate the following expression 
%\begin{equation}
%\mathscr F=
%\frac1{\pi}\PV\left(
%\int_{0}^{\bar \q}
%\chi(\q-\bar \q)\,
%\Pi_F^{(1)}
%\cos n\q\,\d \q
%+
%\int_{\bar \q}^{\pi}
%\chi(\q-\bar \q)\,
%\Pi_F^{(2)}
%\cos n\q\,\d \q
%%}_{\hat\Pi_n^{(2)}}
%\right).
%\end{equation}
%For the aim of simplicity we will do this only in the case $n=0$.
%One has 
%\begin{gather}
%\Pi_F^{(1)}\sim\frac{C_2}{\q-\bar \q}, \qquad \q\to\bar \q-0;
%\\
%\Pi_F^{(2)}\sim\frac{C_2}{\q-\bar \q}, \qquad \q\to\bar \q+0,
%\end{gather}
%where $C_2$ is a non-zero constant. 
%Thus the principal value of the corresponging integral equals
%\begin{equation}
%\mathscr F=
%\frac1{\pi}\left(
%\int_{0}^{\bar \q}
%\chi(\q-\bar \q)\,
%(
%\Pi_F^{(1)}
%+
%\Pi_F^{(0)}
%)
%\cos n\q\,\d \q
%+
%\int_{\bar \q}^{\pi}
%\chi(\q-\bar \q)\,
%(
%\Pi_F^{(2)}
%+
%\Pi_F^{(0)}
%)
%\cos n\q\,\d \q
%%}_{\hat\Pi_n^{(2)}}
%\right).
%\end{equation}
%Taking~${\epsilon}$ in  
%\eqref{FAST-chi-def} to be small enough, 
We asymptotically 
approximate integrand in the neighbourhood of $\q=\bar \q$ using 
Eq.~\eqref{FAST-app-phi},
%\eqref{FAST-psi-approx},
and 
\begin{equation}
P(\q)={\mathrm e}^{-{\eta}\,t}\big(P_{-1}(\q-\bar \q)^{-1}+P_{0}+\dots\big).
\label{FAST-P-expansion}
\end{equation}
Accordingly, one has
\begin{gather}	
R\sim R_{-1}+R_0+\dots,
\label{FAST-R-expansion}
\end{gather}
where
\begin{multline}	
R_{-1}\propto
{
{\mathrm e}^{-{\eta}\,t}}
\left(
\int_{\bar \q-{\epsilon}}^{\bar \q}
%\chi(\q-\bar \q)\,
\frac{1-\cosh\big(\phi_{1/2}|\bar \q-\q|^{1/2}\,t\big)}{\q-\bar \q}
\,\d \q
%+
%\int_{\bar \q-{\epsilon}}^{\bar \q}
%%\chi(\q-\bar \q)\,
%\frac{1-\exp\big(-\phi_{1/2}|\bar \q-\q|^{1/2}\,t\big)}{\q-\bar \q}
%\,\d \q
\right.
\\
\left.
+
\int_{\bar \q}^{\bar \q+{\epsilon}}
%\chi(\q-\bar \q)\,
\frac{1-\cos\big(\phi_{1/2}|\q-\bar \q|^{1/2}\,t\big)}{\q-\bar \q}
\,\d \q
%}_{\hat\Pi_n^{(2)}}
\right)
\\
=
{\mathrm e}^{-{\eta}\,t}
\left(
\int_{-{\epsilon}}^0
%\chi(\q)\,
\frac{1-\cosh\big(\phi_{1/2}|\q|^{1/2}\,t\big)}{\q}
\,\d \q
\right.
\\+
\left.
\int_0^{{\epsilon}}
%\chi(\q)\,
\frac{1-\cos\big(\phi_{1/2}|\q|^{1/2}\,t\big)}{\q}
\,\d \q
\right)
\\
=2
{\mathrm e}^{-{\eta}\,t}
\left(
\Chi(\sqrt{\epsilon} \phi_{1/2}t)
-
\Ci(\sqrt{\epsilon} \phi_{1/2}t)
\right).
\label{FAST-int-cosine-codhine}
\end{multline}
Here $\phi_{1/2}$ is defined by 
\eqref{FAST-app-phi},
$\Ci(\cdot)$ is the integral cosine, $\Chi(\cdot)$ is the integral hyperbolic 
cosine~\cite{abramowitz1972handbook}.
%One has~\cite{?}:
%\begin{gather}
%\Ei(\tau)\sim\frac{\exp(-\tau)\big(1+O(\tau^{-1})\big)}{\tau},\\
%\Chi(\tau)\sim\frac{\exp\tau\big(1+O(\tau^{-1})\big)}{2\tau},\\
%\Ci(\tau)\sim
%\frac{\sin\tau\big(1+O(\tau^{-2})\big)}{\tau}
%-
%\frac{\cos\tau\big(1+O(\tau^{-2})\big)}{\tau^2}.
%\end{gather}
%as $\tau\to\infty$.
Thus, for large $t$ the reminder $R$ is the sum of contributions from the boundary
points $\bar \q\pm {\epsilon}$,
%:
%\begin{equation}
%R\sim
%2C_3
%{\mathrm e}^{-{\eta}\,t}
%\left(
%\Chi(\sqrt{\epsilon} C_1t)
%+O(t^{-1})
%\right)
%=O\left(\frac{\mathrm{e}^{(-\eta+\sqrt{\epsilon} C_1)t}}{\sqrt{\epsilon} t} \right)
%,
%\end{equation}
which must be totally compensated in the sum with the corresponding
contributions from the boundary points for integrals 
$\hat\Pi_n^{(1)(\pm)}$ 
and
$\hat\Pi_n^{(2)(\pm)}$. Hence, to estimate the contribution to
integral 
\eqref{FAST-Pi-with-R}
from 
a neighbourhood of
the point $\q=\bar \q$ we need to take into account the correction terms in
expansions 
\eqref{FAST-app-phi}, \eqref{FAST-P-expansion}.

We introduce the substitution 
\begin{equation}
\phi(\q)=\sqrt Q,
\end{equation}
where $\phi(\q)$ is defined by
Eq.~\eqref{FAST-app-psipm}
(and expansion 
\eqref{FAST-phi-approx}). One has
\begin{equation}
\d \q=\d Q\big(1+O(Q)\big).
\label{FAST-dqQ}
\end{equation}
Analogously, the next term $R_0$ in 
\eqref{FAST-R-expansion} is the sum of contribution of the boundary points
(which again are totally compensated) and a
contribution from $\q=\bar \q$. 
Taking into account
\eqref{FAST-P-expansion}
and 
\eqref{FAST-dqQ},
one can estimate the latter term as follows:
\begin{multline}
R_0(\bar \q)\propto
{{\mathrm e}^{-{\eta}\,t}}
\left(
\int_{-\epsilon}^{0}
\chi(Q)\,
{\cosh\big(C_1|Q|}^{1/2}\,t\big)
\,\d Q
\right.
\\+
\left.
\int_{0}^{{\epsilon}}
\chi(Q)\,
{\cos\big(C_1|Q|^{1/2}\,t\big)}
\,\d Q
%}_{\hat\Pi_n^{(2)}}
\right)=
O\bigg(\frac {\mathrm e^{-\eta t}}{t^2}
\bigg).
\end{multline}
The last formula is obtained using the Erdeliy lemma and Watson lemma in
the case 
$\beta=1,\ \alpha=1/2$ and taking $\bar\epsilon<\epsilon$ in
Eq.~\eqref{FAST-chi-def}.

%On the other hand, the argument of the exponent in the integrand of $R_1$ is
%just the approximation for $t\psi_+(\q)$,
%the integral $R_1$ clearly can be estimated as 
%\begin{equation}
%R_1=o\big(
%\hat\Pi_n^{(1)(+)}
%\big).
%\end{equation}

Thus, the asymptotics of $\hat{\Pi}_n$ equals the sum of the right-hand side of 
Eq.~\eqref{FAST-Pi-App-as-osc}, the right-hand side of
Eq.~\eqref{FAST-Pi1-app}, and the term $\hat\Pi_n^{(0)}$ calculated in
Appendix~\ref{FAST-App-P0} in the particular case $n=0$.

\section{Calculation of the integral $\hat\Pi_0^{(0)}$}
\label{FAST-App-P0}
Since $\bar\epsilon$ 
introduced by Eqs.~\eqref{FAST-Pi-with-R}--\eqref{FAST-reminder}
is an arbitrary positive number, the integral $\hat\Pi_0^{(0)}$ can be
calculated as 
\begin{equation}
\hat\Pi_n^{(0)}
=
\frac1\pi
\PV
%\left(
\int_{0}
%^{\bar \q-{\bar\epsilon}}+
%\int_{\bar \q+{\bar\epsilon}}
^\pi
%\right)
\Pi_F^{(0)}
\cos{n\q}\,\d \q,
\label{FAST-Pi0n-int}
\end{equation}
where symbol $\PV$ means the Cauchy principal value for 
the corresponding improper integral. 

In what follows, we consider only the case $n=0$.
According to 
Eqs.~\eqref{FAST-def-Pi0F}, \eqref{FAST-A-def}
one gets:
\begin{multline}
\hat\Pi_F^{(0)}=
\frac{{\mkE}{\mathrm e}^{-{\eta}\,t}}{2\pi}
\frac {(4\omega_0 \sin \q/2)^2}{(4\omega_0 \sin \q/2)^2-\eta^2}
\\=
\frac{{\mkE}{\mathrm e}^{-{\eta}\,t}}{2\pi}
\left(
1+\frac{\eta^2}{8\omega_0^2-\eta^2-8\omega_0^2\cos \q}
\right).
\label{FAST-Pi0F-exprapp}
\end{multline}
Thus,
\begin{gather}
\hat\Pi_0^{(0)}=
\frac{{\mkE}{\mathrm e}^{-{\eta}\,t}}2\,\big(1+\pi^{-1}\mathscr P\big),
\\
\begin{multlined}	
\mathscr P=\PV
\int_0^\pi
\frac{\eta^2\,\d \q}{8-\eta^2-8\cos \q}
=\eta^2\Big(
\mathscr J(\pi)-\mathscr J(0)
\\+\lim_{\epsilon\to+0}
\big(\mathscr J(\bar \q-\epsilon)-\mathscr J(\bar \q+\epsilon)
\big)
\Big)
\Big|_{a=8-\eta^2,\ b=-8},
\end{multlined}
\end{gather}
where $\mathscr J(\q)$ is the following integral that can be calculated in
the closed form \cite{PBM1}
\begin{gather}
\mathscr J(\q)=\int\frac{\d \q}{a+b\cos \q}=\frac1{\sqrt{b^2-a^2}}\,\ln
\big |
\mathscr Z(\q)
\big |,
\quad
|a|<|b|;
\label{FAST-App-iint}
\\
\mathscr Z(\q)=
\frac
{\sqrt{b^2-a^2}\tan \frac \q2+a+b}
{\sqrt{b^2-a^2}\tan \frac \q2-a-b}.
\end{gather}
One can see that
\begin{equation}
\mathscr J(0)=0,\qquad \mathscr J(\pi)=0,
\end{equation}
whereas at $\q=\bar \q$ the argument $\big|\mathscr Z(\q)\big|$ 
of logarithmic function has indeterminate form
$0/0$. One can obtain the following expansions ($\epsilon\to+0$):
\begin{equation}
\begin{gathered}	
\mathscr Z(\bar \q+\epsilon)
=-\frac
{4\omega_0^2}{\eta\sqrt{16\omega_0^2-\eta^2}}\,\epsilon+O(\epsilon^2),\\
\mathscr Z(\bar \q-\epsilon)
=\frac {4\omega_0^2}{\eta\sqrt{16\omega_0^2-\eta^2}}\,\epsilon+O(\epsilon^2).
\end{gathered}
\label{FAST-App-iint-barq}
\end{equation}
Now it follows from Eqs.~\eqref{FAST-App-iint-barq}, \eqref{FAST-App-iint}
that
\begin{equation}
\lim_{\epsilon\to+0}
\big(\mathscr J(\bar \q-\epsilon)-\mathscr J(\bar \q+\epsilon)
\big)=0,
\end{equation}
and 
\begin{equation}
\hat\Pi_0^{(0)}=
\frac{{\mkE}{\mathrm e}^{-{\eta}\,t}}2.
\end{equation}

%%%\section{The proof for uniform convergence of $R_0(\epsilon,t)$ to zero as
%%%$\epsilon\to+0$}
%%%\label{FAST-App-uni}

%\end{subappendices}
\vspace*{1cm}
\bibliographystyle{apsrev4-2}
%\clearpage
\bibliography{math,thermo,serge-gost}